\documentclass[11pt]{article}
\usepackage[dvips]{graphicx}
\usepackage{bm}
\usepackage{amsmath}
\usepackage{amssymb}
\usepackage{float}
\usepackage{url}
\renewcommand{\baselinestretch}{1.1}
\renewcommand{\thepage}{}
\makeatletter
\@addtoreset{equation}{section}
\renewcommand{\theequation}{\thesection.\@arabic\c@equation}
\makeatother
\renewcommand{\thefootnote}{\fnsymbol{footnote}}
\begin{document}
\begin{titlepage}
\title{
\vspace*{-4ex}
\hfill{}\\
\hfill
\begin{minipage}{3.5cm}
\end{minipage}\\
%\vspace{4ex}
\bf Numerical twist-even SU(1,1)-singlet solutions
 in open string field theory around the identity-based solution
\vspace{3ex}
}

%\normalsize
\author{
Isao~{\sc Kishimoto}$^{1}$\footnote{ikishimo@rs.socu.ac.jp}
~and~ 
Tomohiko~{\sc Takahashi}$^{2}$\footnote{tomo@asuka.phys.nara-wu.ac.jp}
\\
\vspace{0.5ex}\\
%\vspace{2ex}\\
\\
$^{1}${\it Center for Liberal Arts and Sciences, Sanyo-Onoda City University,}\\
{\it Daigakudori 1-1-1, Sanyo-Onoda Yamaguchi 756-0884, Japan}
\vspace{1ex}
\\
$^{2}${\it Department of Physics, Nara Women's University,}\\
{\it Nara 630-8506, Japan}
\vspace{2ex}
}

%\date{\today}
\date{}
\maketitle
%\tableofcontents

%
\vspace{7ex}

\begin{abstract}
\normalsize
Using the level truncation method, we construct numerical solutions,
which are twist even and SU(1,1) singlet, in the theory around the
Takahashi-Tanimoto identity-based solution (TT solution) with a real
parameter $a$ in the framework of bosonic open string field theory.  We
find solutions corresponding to ``double brane" and ``ghost brane"
solutions which were constructed by Kudrna and Schnabl in the
conventional theory around the perturbative vacuum.  Our solutions show
somewhat similar $a$-dependence to tachyon vacuum and single brane
solutions, which we found in the earlier works.  In this sense, we might
be able to expect that they are consistent with the conventional
interpretation of $a$-dependence of the TT solution.  We observe that
numerical complex solutions at low levels become real ones at higher
levels for some region of the parameter $a$.
However, these real solutions do not so improve interpretation 
for double brane.
\end{abstract}

\vspace{5ex}

\end{titlepage}

%%%%%%%%%%%%%%%%%%%%%%%%%%%%%%%%%%%%%%%%%%%%%%%%%%%%%%
\renewcommand{\thepage}{\arabic{page}}
\renewcommand{\thefootnote}{\arabic{footnote}}
\setcounter{page}{1}
\setcounter{footnote}{0}
%\baselineskip=19pt plus 0.2pt minus 0.1pt
%%%%%%%%%%%%%%%%%%%%%%%%%%%%%%%%%%%%%%%%%%%%%%%%%%%%%%
%
\renewcommand{\baselinestretch}{1}
\tableofcontents
\renewcommand{\baselinestretch}{1.1}
%%%%%

\newpage
\section{Introduction}

In bosonic open string field theory, there is a well-known numerical
solution: the tachyon vacuum solution in Siegel gauge.  It was found by
Sen and Zwiebach \cite{Sen:1999nx} using the level truncation method and
then higher level calculations were performed for the solution
\cite{Moeller:2000xv, Gaiotto:2002wy, Kishimoto:2011zza,
Kudrna:2018mxa}.  It is known that efficient and consistent truncation
for it can be obtained by restricting the space of string fields to that
spanned by twist even and SU(1,1) singlet states.  With the same
restriction, but relaxing the reality condition, Kudrna and Schnabl
constructed two interesting solutions \cite{Kudrna:2018mxa}, which might
be interpreted as double brane and ghost brane, respectively.

Numerical solutions mentioned above can be constructed using Newton's
method by choosing appropriate initial configurations so that iterative
calculations converge.  The tachyon vacuum solution can be uniquely
obtained from a real solution at the level 0, namely, the lowest
truncation level.  The ``double brane" and ``ghost brane" solutions can
be obtained from one of complex solutions, which do not satisfy the
reality condition for string fields, at the truncation level 2 and 4,
respectively.

On the other hand, the Takahashi-Tanimoto (TT) solution
\cite{Takahashi:2002ez} is based on the identity string field and
therefore direct evaluation of its energy was difficult.  Alternatively,
in the theory around the TT solution, the BRST cohomology was studied
\cite{Kishimoto:2002xi} and numerical solutions in Siegel gauge were
investigated \cite{Takahashi:2003ppa, Kishimoto:2009nd,
Kishimoto:2009hc}.  The TT solution has a real parameter $a$ such as
$a\ge -1/2$ and it is expected that it represents the tachyon vacuum at
$a=-1/2$ and is pure gauge for $a>-1/2$. It was supported by analysis
of cohomology in \cite{Kishimoto:2002xi} and evaluation of energy of the
tachyon vacuum solution in Siegel gauge \cite{Takahashi:2003ppa} in the
theory around the TT solution ($\Psi_a^{\rm TT}$).  Furthermore,
numerical solution for unstable vacuum, which corresponds to the
perturbative vacuum or single brane, was found in
\cite{Kishimoto:2009nd} in the theory around
$\Psi_a^{\rm TT}$ for $a\simeq -1/2$. \footnote{The energy of
$\Psi_a^{\rm TT}$ is calculated analytically in
\cite{Ishibashi:2014mua,Kishimoto:2014lua} and it is confirmed that the
TT solution is the tachyon vacuum at $a=-1/2$ and is pure gauge for $a>-1/2$.}

In this paper, motivated by ``double brane" and ``ghost brane" solutions
found in \cite{Kudrna:2018mxa}, we construct numerical twist-even
SU(1,1)-singlet solutions, which correspond to them, in the theory
around $\Psi_a^{\rm TT}$ for $a\ge -1/2$.  They show somewhat similar
$a$-dependence to the tachyon vacuum and single brane solutions.
Namely, with increasing level, energy of them seems to approach a
constant $E$ for $a>-1/2$ and another value $E^{\prime}$ for $a\simeq
-1/2$ ($E^{\prime}>E$).  However, the numerical behaviors of them are
not so clear.  Actually, the values of energy are complex in general
because these numerical solutions are obtained from one of complex
solutions in the truncation level 2 and 4.  Roughly, the imaginary part
of the energy for them seems to approach zero with increasing level.
Particularly, we find that these solutions in the theory around
$\Psi_a^{\rm TT}$ for some region of $a$ satisfy the reality condition at
higher levels although they start from complex solutions at the level 2
or 4.
 
Here, we have performed numerical calculations up to the truncation
level 22, and the values of the energy and the gauge invariant
observable\footnote{ It is also called gauge invariant overlap or
Ellwood invariant in the literatures.}  for ``double brane" and `` ghost
brane" solutions might (not) represent the number of branes literally.
Further investigation is necessary for a definite interpretation.
 
This paper is organized as follows.  In \S \ref{sec:Newton}, we will
explain our strategy of numerical calculations and conventions briefly.
In \S \ref{sec:Dsol} and \S \ref{sec:Gsol}, we will show our numerical
results: plots of energy, gauge invariant observables and so on, for
``double brane" solution and ``ghost brane" solution, respectively.  In
\S \ref{sec:remarks}, we will give some remarks on our calculations.
Moreover, in appendix \ref{sec:App}, we will give some explicit
numerical data for the solutions at $a=-1/2$ including evaluations of
quadratic identities.
 
\section{
 Procedure for constructing numerical solutions
 \label{sec:Newton}
}
 
In this section, we briefly explain procedure to construct numerical
solutions and we define quantities to evaluate for the solutions.
Further technical details can be found in \cite{Kudrna:2018mxa}.
 
The TT solution $\Psi^{\rm TT}_a$ \cite{Takahashi:2002ez},
with a  real parameter $a$ such as $a\ge -1/2$,
is one of identity-based solutions to the equation of motion $Q_{\rm
 B}\Psi+\Psi\ast\Psi=0$ in open string field theory, whose action is
 $S[\Psi]=-\frac{1}{g^2}\left( \frac{1}{2}\langle\Psi,Q_{\rm
 B}\Psi\rangle+\frac{1}{3}\langle\Psi,\Psi\ast\Psi\rangle \right)$,
where $Q_{\rm B}$ is the conventional BRST operator.  Around the TT
solution $\Psi^{\rm TT}_a$, we define the action $S_a[\Phi]$
as\footnote{ We take the case of $l=1$ for the TT solution, where an
integer $l$ was introduced in \cite{Kishimoto:2002xi}.  Similar
computations can be performed for $l=2,3,4,\cdots$, but $l=1$ is the
most fundamental in the context of the level truncation in the sense
that mixing of the levels in the kinetic term is minimum in Siegel
gauge.  }
\begin{align}
S_a[\Phi]&=S[\Psi^{\rm TT}_a+\Phi]-S[\Psi^{\rm TT}_a]
=-\dfrac{1}{g^2}
\left(\dfrac{1}{2}\langle\Phi,Q^{\prime}\Phi\rangle+\frac{1}{3}\langle\Phi,\Phi\ast\Phi\rangle\right),
\label{eq:S_aPhi}
\\
Q^{\prime}&=(1+a)Q_{\rm B}+\dfrac{a}{2}(Q_2+Q_{-2})+4aZ(a)c_0-2aZ(a)^2(c_2+c_{-2})
\nonumber\\
&\quad+2a(1-Z(a)^2)\sum_{n=2}^{\infty}(-Z(a))^{n-1}(c_{2n}+c_{-2n}),\\
Z(a)&=\dfrac{1+a-\sqrt{1+2a}}{a},
\end{align}
where $Q_n$ is a mode of the BRST current.  In the case $a=0$,
$\Psi_a^{\rm TT}$ becomes zero, namely $\Psi_{a=0}^{\rm TT}=0$, and hence
$Q^{\prime}|_{a=0}=Q_{\rm B}$.  The equation of motion of
(\ref{eq:S_aPhi}) is
\begin{align}
 Q^{\prime}\Phi+\Phi\ast\Phi=0
 \label{eq:EOMQp}
\end{align}
 and it is projected to
\begin{align}
&L(a)\Phi+b_0(\Phi\ast\Phi)=0,
\label{eq:EOM_La}
\\
&L(a)=(1+a)(L_0^{\rm mat}+L_0^{\prime {\rm gh}}-1)
+\dfrac{a}{2}(L_2^{\rm mat}+L_2^{\prime{\rm gh}}
+L_{-2}^{\rm mat}+L_{-2}^{\prime{\rm gh}})
+4(1+a-\sqrt{1+2a}),
\end{align}
using the Siegel gauge condition: $b_0\Phi=0$.  Here, $L_n^{\rm mat}$ is
the Virasoro generator in the matter sector with the central charge $26$
and $L_n^{\prime{\rm gh}}$ is a twisted Virasoro generator in the ghost
sector with the central charge $-2$, which is given by
 \begin{align}
 L_n^{\prime{\rm gh}}&=\sum_{m=-\infty}^{\infty}(n-m):b_mc_{n-m}:
 \end{align}
in terms of the $bc$-ghost modes.  In order to solve a nonlinear
equation (\ref{eq:EOM_La}), we use Newton's method as follows.  Firstly,
we take an initial string field $\Phi^{(0)}$, then we solve a linearized
equation:
\begin{align}
L(a)\Phi^{(n+1)}+b_0(\Phi^{(n)}\ast\Phi^{(n+1)})+b_0(\Phi^{(n+1)}\ast\Phi^{(n)})
=b_0(\Phi^{(n)}\ast\Phi^{(n)})
\label{eq:LaPhi_iteeq}
\end{align}
for $n=0,1,2,\cdots$, iteratively.  If $\Phi^{(n)}$ converges to a
string field with $n\to \infty$, $\Phi^{(\infty)}$ is a solution to
(\ref{eq:EOM_La}).  Actually, we stop the calculation when
$\|\Phi^{(n+1)}-\Phi^{(n)}\|/\|\Phi^{(n)}\|<\varepsilon$ for a
sufficiently small positive constant $\varepsilon$, where $\|~\cdot~\|$
is a norm and we regard $\Phi^{(n+1)}$ as an approximate solution.  

We adopt the level truncation method to take a finite number of component
fields from a string field $\Phi$ for numerical calculation.  As a
consistent level $L$ truncation, where $L$ is the eigenvalue of
$L_0^{\rm mat}+L_0^{\prime{\rm gh}}$, we expand a string field of the
ghost number one with a basis which consists of twist even and SU(1,1)
singlet states of the form:
\begin{align}
&L_{-n_m}^{\rm mat}\cdots L_{-n_1}^{\rm mat}L_{-l_g}^{\prime{\rm gh}}
\cdots L_{-l_1}^{\prime{\rm gh}}c_1|0\rangle,
\label{eq:LT-basis}
\\
&n_m\ge \cdots \ge n_1\ge 2,\qquad l_g\ge \cdots \ge l_1\ge 2,\qquad
\sum_{k=1}^mn_k+\sum_{k=1}^gl_k=\ell,\qquad (\ell = 0, 2, 4,\cdots, L).
\end{align}
The truncation level $L$ is an even integer from the twist even
condition and $|0\rangle$ is the conformal vacuum.  We denote a state of
the above form as $\psi_i$ and we expand $\Phi$ as
\begin{align}
&\Phi=\sum_{i=1}^{N_L}t_i\psi_i,
\end{align}
with coefficients $t_i$, where $N_L$ is the dimension of the truncated
state space up to level $L$, namely, $N_L=1,3,8,21,51,117,\cdots$ for
$L=0,2,4,6,8,10,\cdots$, respectively.  We take $t_i$ as a complex
constant in general although it should be real from the reality
condition of string fields.  It is necessary to construct ``double
brane" and ``ghost brane" solutions as we will see later.

With the above level truncation of string fields, we take a BPZ inner
product of $c_0\psi_i$ and (\ref{eq:LaPhi_iteeq}) and we obtain
simultaneous equations:
\begin{align}
\sum_{j=1}^{N_L}\left((L(a))_{ij}+2\sum_{k=1}^{N_L}V_{ijk}t_k^{(n)}\right)t_j^{(n+1)}&=
\sum_{j,k=1}^{N_L}V_{ijk}t_k^{(n)}t_j^{(n)},
\label{eq:simueqs}
\end{align}
where 
\begin{align}
&(L(a))_{ij}=\langle \psi_i,c_0L(a)\psi_j\rangle,
&&V_{ijk}=\langle \psi_i,\psi_j\ast\psi_k\rangle,
\end{align}
and use has been made of $V_{ijk}=V_{ikj}$ thanks to the twist even
condition.  With appropriate initial values
$\{t_i^{(0)}\}_{i=1,2,\cdots,N_L}$ $(t_i^{(0)}\in {\mathbb C})$, we
solve (\ref{eq:simueqs}) iteratively for $n=0,1,2,\cdots$.  We stop the
calculation if $\|{\bm t}^{(n+1)}-{\bm t}^{(n)}\|/\|{\bm t}^{(n)}\|<
\varepsilon$ with the Euclidean norm for a sufficiently small positive
$\varepsilon$ and we regard
\begin{align}
\Phi_a&=\sum_{i=1}^{N_L}\tilde t_i\psi_i.
\end{align}
($\tilde t_i=t_i^{(n+1)}$) as a numerical solution to
(\ref{eq:EOM_La}).\footnote{ In our actual calculation in \S
\ref{sec:Dsol} and \S \ref{sec:Gsol}, we took $\varepsilon=5\times
10^{-14}$ in our C++ code with the long double format. 
 If $\|{\bm t}^{(n+1)}-{\bm t}^{(n)}\|/\|{\bm t}^{(n)}\|\ge
\varepsilon$ for $n+1=15$, we regarded it as not converging.  }

For a numerical solution $\Phi_a$, we compute the energy $E[\Phi_a]$,
which is given by the action (\ref{eq:S_aPhi}) as
\begin{align}
E[\Phi_a]&=1-2\pi^2 g^2 S_a[\Phi_a]=1+\dfrac{\pi^2}{3}\sum_{i,j=1}^{N_L}
(L(a))_{ij}{\tilde t}_i{\tilde t}_j,
\label{eq:EPhia}
\end{align}
where we have used the equation of motion for $\tilde{\bm t}$ :
\begin{align}
\sum_{j=1}^{N_L}(L(a))_{ij}{\tilde t}_j
+\sum_{j,k=1}^{N_L}V_{ijk}{\tilde t}_k{\tilde t}_j=0.
\label{eq:eomtilt}
\end{align}
$E[\Phi_a]$ (\ref{eq:EPhia}) is normalized in the same way as
\cite{Kudrna:2018mxa}.  In the case $a=0$, or in the theory around the
perturbative vacuum, $E[0]|_{a=0}=1$ for the single
brane solution $0$, which is the perturbative vacuum,
and $E[\Psi^{\rm T}]|_{a=0}=0$ for the tachyon vacuum solution
$\Psi^{\rm T}$.

We evaluate the gauge invariant observable $E_0[\Phi_a]$ for $\Phi_a$ :
\begin{align}
E_0[\Phi_a]=1-2\pi\langle I|V|\Phi_a\rangle
\label{eq:E0Phia}
\end{align}
where $V$ is given by $c(i)c(-i){\mathcal V}(i,-i)$ and ${\mathcal
V}(z,\bar z)$ is matter primary with conformal weight $(1,1)$ and is
normalized as
\begin{align}
\langle I|V\, c_1|0\rangle=\dfrac{1}{4}.
\end{align}
We note that (\ref{eq:E0Phia}) satisfies $E_0[\Psi_{\rm Sch}]=0$
\cite{Kawano:2008ry, Ellwood:2008jh} for Schnabl's analytic solution for
tachyon condensation $\Psi_{\rm Sch}$ \cite{Schnabl:2005gv}.

%%%%%%
For the tachyon vacuum solution $\Phi^{\rm T}_a$
and the perturbative vacuum (or single brane) solution $\Phi^{\rm S}_a$,
in the theory around the TT solution $\Psi_a^{\rm TT}$  ($a\ge -1/2$),
it has been shown that $E$ (\ref{eq:EPhia}) and $E_0$ (\ref{eq:E0Phia}) behave as
\begin{align}
&E[\Phi^{\rm T}_a]\to 
\begin{cases}
0&(a>-1/2)\\
1&(a=-1/2)
\end{cases},
&&
E_0[\Phi^{\rm T}_a]\to 
\begin{cases}
0&(a>-1/2)\\
1&(a=-1/2)
\end{cases},
\label{eq:EE0PhiT}
\\
&E[\Phi^{\rm S}_a]\to 
\begin{cases}
1&(a>-1/2)\\
2&(a=-1/2)
\end{cases},
&&
E_0[\Phi^{\rm S}_a]\to 
\begin{cases}
1&(a>-1/2)\\
2&(a=-1/2)
\end{cases},
\label{eq:EE0PhiS}
\end{align}
numerically in the large truncation level limit  $L\to \infty$,
where $\Phi^{\rm T}_{a=-1/2}\to 0$ and $\Phi^{\rm S}_{a>-1/2}\to 0$ \cite{Kishimoto:2009nd}.
 %%%%%%%
 
As a consistency of the equation of motion (\ref{eq:EOMQp}) for
numerical solutions to (\ref{eq:EOM_La}), we evaluate
\begin{align}
{}|\Delta_S[\Phi_a]|&=\left|
\langle 0|c_{-1}b_{2}c_0|Q^{\prime}\Phi_a+\Phi_a\ast\Phi_a\rangle
\right|
\label{eq:absDelta_S}
\end{align}
after \cite{Kudrna:2018mxa}. It is the lowest level verification of BRST
invariance for Siegel gauge solutions in the context of
\cite{Hata:2000bj}.  Furthermore, we evaluate
\begin{align}
\text{Im}/\text{Re}\,[\Phi_a]=\dfrac{\|{\rm Im}\,\tilde{\bm t}\|}{\|{\rm Re}\,\tilde{\bm t}\|}
\label{eq:Im/RePhi_a}
\end{align}
for reality of numerical solutions, which is given by the ratio of the
Euclidean norm of imaginary and real part of $\tilde{\bm t}$.
 
\section{ ``Double brane" solution
 \label{sec:Dsol} }

In the theory around the perturbative vacuum, which is the case $a=0$ in
(\ref{eq:S_aPhi}), a ``double brane" solution $\Phi^{\rm D}_{a=0}$ is
obtained from one of complex solutions at the truncation level $2$.
Taking a solution $\Phi^{\rm D}_{a=0}$ at level $L$ as an initial string
field, we can obtain $\Phi^{\rm D}_{a=0}$ at level $L+2$ by Newton's
method as explained in \S \ref{sec:Newton}.  Such a solution $\Phi^{\rm
D}_{a=0}$ coincides with the ``double brane" solution in
\cite{Kudrna:2018mxa}.

In the case $a\ne 0$, we adopt a strategy to construct solutions
$\Phi^{\rm D}_a$ corresponding to ``double brane" as follows:
\begin{enumerate} 
\item At level $2$, we construct a solution $\Phi^{\rm D}_{a\mp
      \epsilon}$ using Newton's method
with an initial string field $\Phi^{\rm D}_{a}$.
We repeat this calculation with $\epsilon=0.001$ 
up to $\Phi^{\rm D}_{a=\mp 1/2}$ from the starting point $\Phi^{\rm
      D}_{a=0}$.

 \item At level $L+2$, we construct a solution $\Phi^{\rm D}_a$ using
      Newton's method with an initial string field $\Phi^{\rm D}_{a}$ at
      level $L$. We repeat this calculation up to level $22$.
\end{enumerate}
Namely, for a fixed value of $a$, which is one of
$a=-0.5,-0.499,-0.498,\cdots,0.499,0.5$, we constructed higher level
solutions $\Phi^{\rm D}_a$ from level 2 up to $22$, level by level.
  
Then, we have obtained numerical solutions $\Phi^{\rm D}_a$, except for
$\Phi^{\rm D}_{a=-0.468}$ at levels $20$ and $22$.  In the case
$a=-0.468$, Newton's method for getting a solution at level 20 did not
converge.

\subsection{Energy
\label{sec:EDsol}}

\begin{figure}[htbp]
 \centering
 \includegraphics[height=6.5cm]{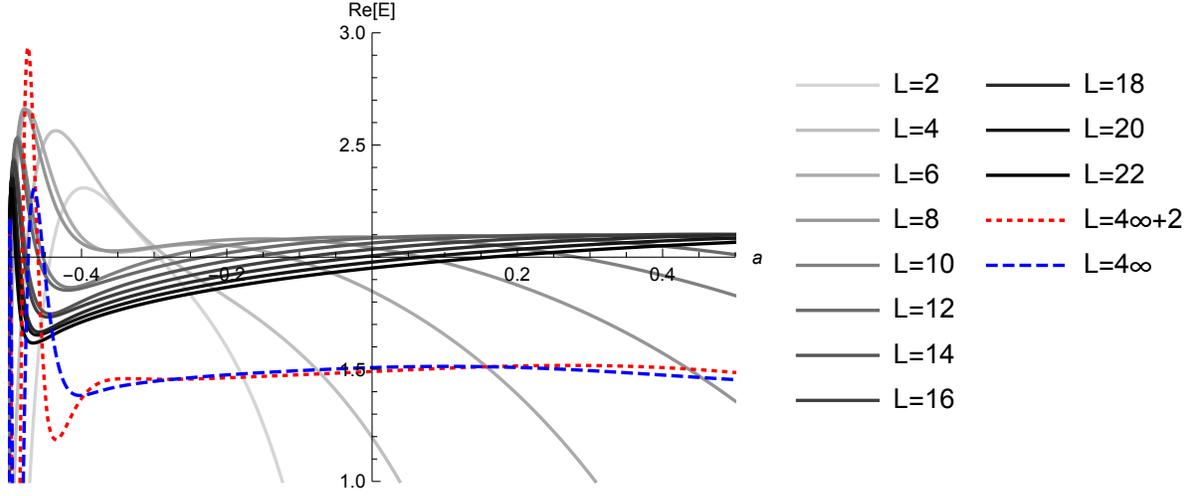}
 \caption{
 Plots of the real part of the energy $E$ (\ref{eq:EPhia}) for the  ``double brane" solution $\Phi_a^{\rm D}$
 at the truncation level $L$.
 The dotted and dashed lines are extrapolations to $L=4k+2$ ($k\to \infty$) and $L=4k$ ($k\to \infty$),
 respectively.
 The horizontal axis denotes the value of the parameter $a$ at ${\rm Re}\,E=2$.
 \label{fig:re_e_dsol}
 }
 \end{figure}

\begin{figure}[htbp]
  \centering \includegraphics[height=6.5cm]{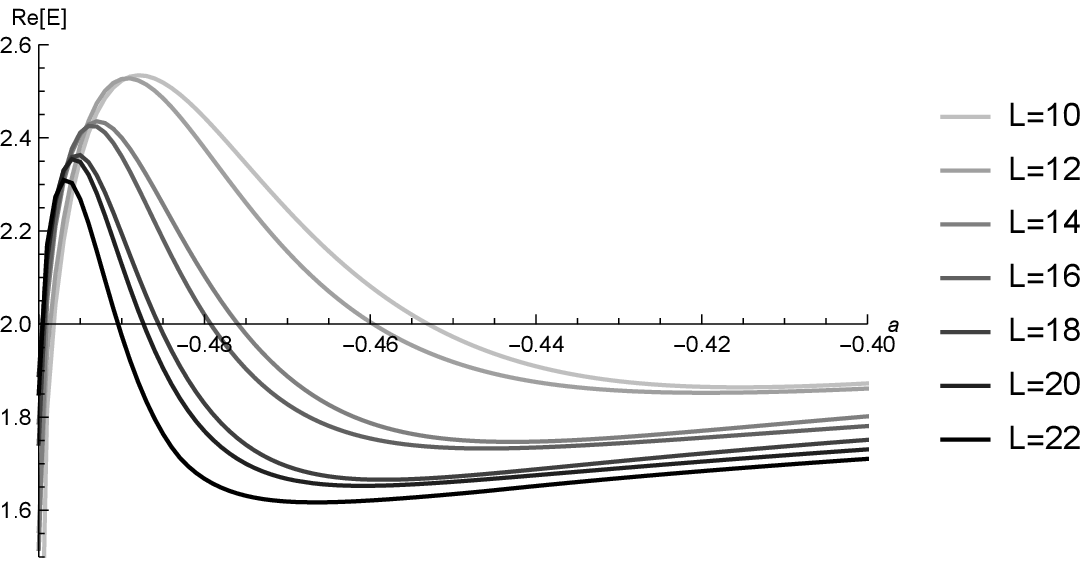} \caption{
 Plots of the real part of the energy $E$ (\ref{eq:EPhia}) for the
 ``double brane" solution $\Phi_a^{\rm D}$ such as $a\gtrsim -1/2$ at
 the truncation level $L=10,12,\cdots,22$.  The horizontal axis denotes
 the value of the parameter $a$ at ${\rm Re}\,E=2$.  The vertical axis
 stands at $a=-1/2$.  
\label{fig:re_e_dsol10} }
 \end{figure}

\begin{figure}[htbp]
 \centering
 \includegraphics[height=6.5cm]{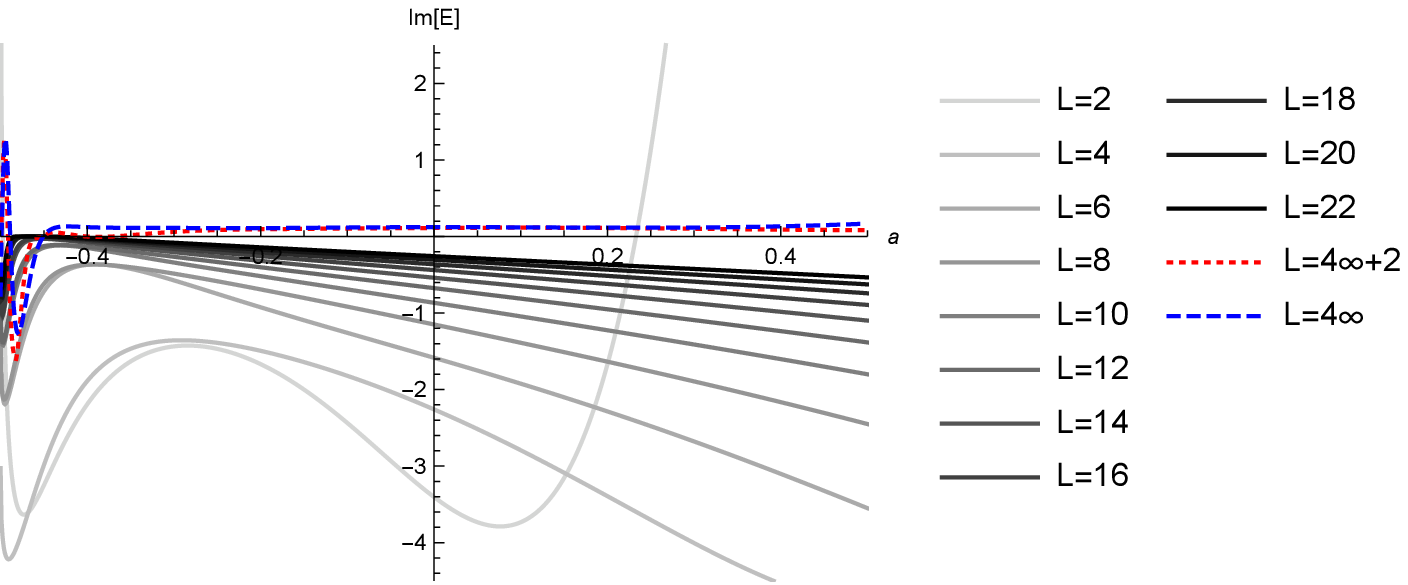}
 \caption{
 Plots of the imaginary part of the energy $E$ (\ref{eq:EPhia}) 
 for the  ``double brane" solution $\Phi_a^{\rm D}$
 at the truncation level $L$.
 The dotted and dashed lines are extrapolations to $L=4k+2$ ($k\to \infty$) and $L=4k$ ($k\to \infty$),
 respectively.
 The horizontal axis denotes the value of the parameter $a$.
 \label{fig:im_e_dsol}
 }
 \end{figure}

Figs.~\ref{fig:re_e_dsol}, \ref{fig:re_e_dsol10}, and \ref{fig:im_e_dsol}
show plots of the energy $E$ (\ref{eq:EPhia}) for the ``double brane"
solution $\Phi_a^{\rm D}$ at the truncation level $L$.  We joined 
adjacent calculated data points with line segments for each
level.\footnote{ Plots of other figures in this paper are in the same
manner.  } In Fig.~\ref{fig:re_e_dsol}, the dotted and dashed lines are
extrapolations to $L=4k+2$ ($k\to \infty$) and $L=4k$ ($k\to \infty$),
respectively.  We have used a polynomial of $1/L$ as a fitting
function for each value of the parameter $a$.  We have chosen its degree
as the number of data minus one.\footnote{ Extrapolations in other
figures in this paper have been obtained in the same manner.  } As might
be expected from Fig.~\ref{fig:re_e_dsol10}, extrapolations by a
polynomial fit worked well by dividing data in two groups: $L\equiv
2\mod 4$ and $L\equiv 0 \mod 4$.

As in Fig.~\ref{fig:re_e_dsol}, the real part of energy ${\rm
Re}\,E[\Phi_a^{\rm D}]$ approaches a constant, which is greater than
one, for $a>-1/2$ with increasing level and there is a maximum at
$a=a_{\rm M}\sim -1/2$ for each $L$.  As in Fig~\ref{fig:re_e_dsol10},
the value of $a_{\rm M}$ approaches $-1/2$ with increasing level, where
the maximum value is greater than two.  The extrapolation values of
${\rm Re}\,E[\Phi_a^{\rm D}]$ in Fig.~\ref{fig:re_e_dsol} are close to
$1.5$ for $a>-1/2$.

As in Fig.~\ref{fig:im_e_dsol}, the imaginary part of energy ${\rm
Im}\,E[\Phi_a^{\rm D}]$ approaches zero for $a>-1/2$ with increasing
level and there is a minimum at $a=a_{\rm m}\sim -1/2$ for each $L$.
The extrapolation values of ${\rm Im}\,E[\Phi_a^{\rm D}]$ in
Fig.~\ref{fig:im_e_dsol} are close to zero for $a>-1/2$.  We observed
that $a_{\rm m}$ approaches $-1/2$ and the minimum value approaches zero
with increasing level although extrapolation values are unstable near
$a=-1/2$.
 
{}From the above observation and analogy with tachyon vacuum and single brane solutions in 
(\ref{eq:EE0PhiT}) and (\ref{eq:EE0PhiS}),
we could expect that  the energy (\ref{eq:EPhia}) 
for the ``double brane" solution $\Phi_a^{\rm D}$ behaves as
\begin{align}
&E[\Phi_a^{\rm D}]\to 
\begin{cases}
E_2 &(a>-1/2)\\
E_2^{\prime} &(a=-1/2)
\end{cases}
&&(L\to \infty),
\label{eq:EaPsiDE2E2p}
\end{align}
where $E_2$ and $E_2^{\prime}$ seem to be real constants and satisfy $1<E_2<E_2^{\prime}$.
If $E_2=2$ and $E_2^{\prime}=3$,  $\Phi_a^{\rm D}$ could be interpreted 
as double brane solution for $a\ge -1/2$ literally.
However, it seems that $E_2\sim 1.5$ and $2<E_2^{\prime}<3$ from Figs.~\ref{fig:re_e_dsol}
and \ref{fig:re_e_dsol10}.
If $E_2^{\prime}-E_2=1$, the solution  $\Phi_a^{\rm D}$  for $a\ge -1/2$
is consistent with the interpretation that the TT solution $\Psi_a^{\rm TT}$
represents the tachyon vacuum at $a=-1/2$ and pure gauge solution for $a>-1/2$,
which implies $S[\Psi_{a=-1/2}^{\rm TT}]=1/(2\pi^2 g^2)$ and
$S[\Psi_{a>-1/2}^{\rm TT}]=0$.

\subsection{Gauge invariant observable
\label{sec:E0Dsol}}

\begin{figure}[htbp]
 \centering 
\includegraphics[height=6.5cm]{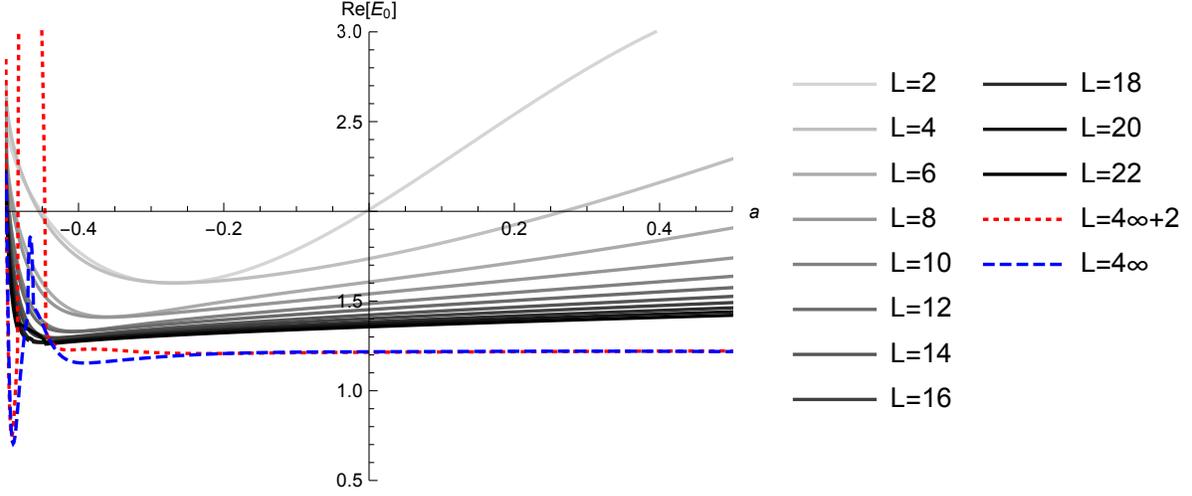} 
\caption{
 Plots of the real part of the gauge invariant observable $E_0$
 (\ref{eq:E0Phia}) for the ``double brane" solution $\Phi_a^{\rm D}$ at
 the truncation level $L$.  The dotted and dashed lines are
 extrapolations to $L=4k+2$ ($k\to \infty$) and $L=4k$ ($k\to \infty$),
 respectively.  The horizontal axis denotes the value of the parameter
 $a$ at ${\rm Re}\,E_0=2$.  \label{fig:re_e0_dsol} }
\end{figure}

\begin{figure}[htbp]
\centering
\includegraphics[height=6.5cm]{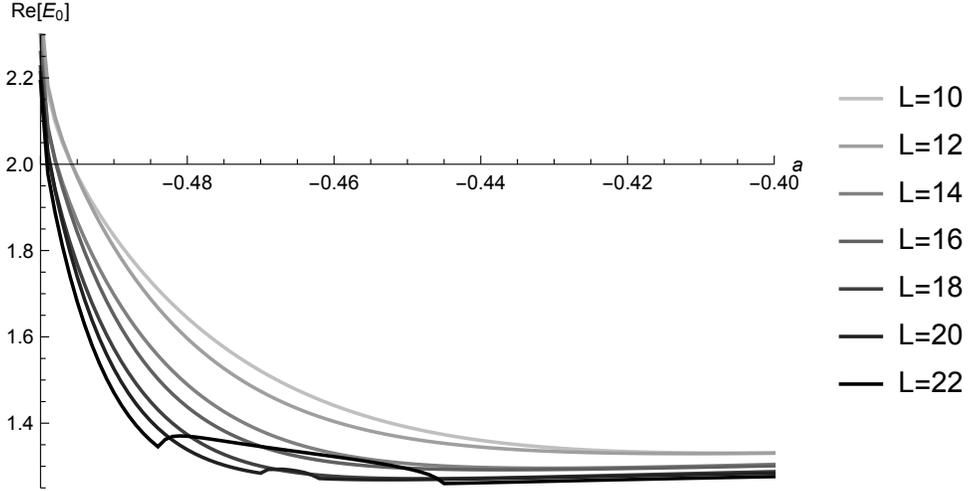} 
\caption{ Plots of the
 real part of the gauge invariant observable $E_0$ (\ref{eq:E0Phia}) for
 the ``double brane" solution $\Phi_a^{\rm D}$ such as $a\gtrsim -1/2$
 at the truncation level $L=10,12,\cdots,22$.  The horizontal axis
 denotes the value of the parameter $a$ at ${\rm Re}\,E_0=2$.  The
 vertical axis stands at $a=-1/2$.  \label{fig:re_e0_dsol10} }
\end{figure}

\begin{figure}[htbp]
 \centering
 \includegraphics[height=6.5cm]{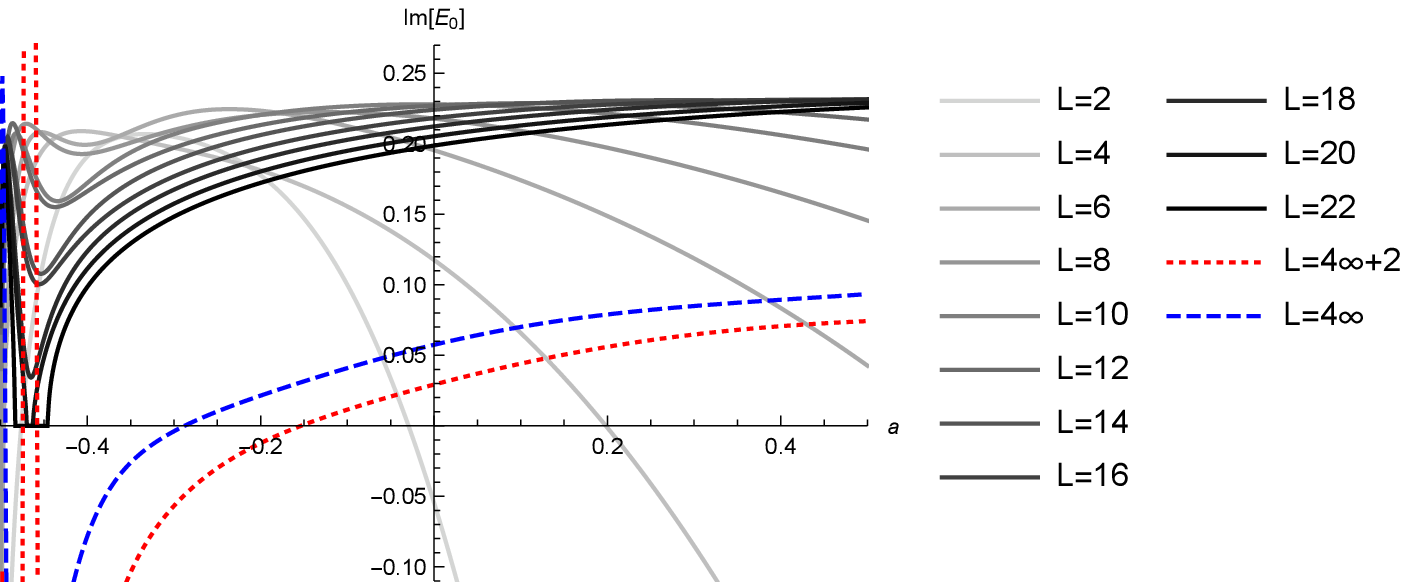}
\caption{
 Plots of the imaginary part of the gauge invariant observable $E_0$
 (\ref{eq:E0Phia}) for the ``double brane" solution $\Phi_a^{\rm D}$ at
 the truncation level $L$.  The dotted and dashed lines are
 extrapolations to $L=4k+2$ ($k\to \infty$) and $L=4k$ ($k\to \infty$),
 respectively.  The horizontal axis denotes the value of the parameter
 $a$.  
\label{fig:im_e0_dsol} }
\end{figure}

Figs.~\ref{fig:re_e0_dsol}, \ref{fig:re_e0_dsol10}, and
 \ref{fig:im_e0_dsol} show plots of the gauge invariant observable $E_0$
 (\ref{eq:E0Phia}) for the ``double brane" solution $\Phi_a^{\rm D}$ at
 the truncation level $L$.
 
As in Fig.~\ref{fig:re_e0_dsol}, the real part ${\rm
Re}\,E_0[\Phi_a^{\rm D}]$ approaches a constant, which is greater than
one, for $a>-1/2$ with increasing level.  As in
Fig.~\ref{fig:re_e0_dsol10} (and Table~\ref{tab:a=m1o2D} in appendix~\ref{sec:App}), 
${\rm Re}\,E_0[\Phi_a^{\rm D}]>2$ for $a\to -1/2+0$.  The extrapolation values
of ${\rm Re}\,E_0[\Phi_a^{\rm D}]$ in Fig.~\ref{fig:re_e0_dsol} are
close to $1.2$ for $a>-1/2$ although they become unstable near $a=-1/2$.
This instability or error should be caused by irregular behavior around
$a=-0.465$ at $L=20,22$ in Fig.~\ref{fig:re_e0_dsol10}.
 
As in Fig.~\ref{fig:im_e0_dsol}, the imaginary part ${\rm
Im}\,E_0[\Phi_a^{\rm D}]$ approaches zero for $a>-1/2$ with increasing
level.  The extrapolation values are around zero for $a>-1/2$ although
they are unstable near $a=-1/2$.  Actually, we have found that ${\rm
Im}\,E_0[\Phi_a^{\rm D}]=0$ around $a=-0.465$ for $L=20,22$.
 
From the above, in a similar way to (\ref{eq:EaPsiDE2E2p}), 
we could expect that the gauge invariant observable (\ref{eq:E0Phia})
behaves as 
\begin{align}
E_0[\Phi_a^{\rm D}]\to 
\begin{cases}
\tilde E_2 &(a>-1/2)\\
\tilde E_2^{\prime} &(a=-1/2)
\end{cases}
&&(L\to \infty),
\label{eq:E0PhiaDE2E2p}
\end{align}
where $\tilde E_2$ and $\tilde E_2^{\prime}$ seem to be real constants
and satisfy $1<\tilde E_2<\tilde E_2^{\prime}$.  If $\tilde E_2=2$ and
$\tilde E_2^{\prime}=3$, $\Phi_a^{\rm D}$ could be interpreted as double
brane solution for $a\ge -1/2$ because the value of (\ref{eq:E0Phia})
corresponds to the energy \cite{Baba:2012cs}.  However, it seems that
$\tilde E_2\sim 1.2$ and $2\lesssim \tilde E_2^{\prime} < 3$ from
Figs.~\ref{fig:re_e0_dsol} and \ref{fig:re_e0_dsol10}.  If $\tilde
E_2^{\prime}-\tilde E_2=1$, the solution $\Phi_a^{\rm D}$ for $a>-1/2$
is consistent with the TT solution in the same sense as the energy.
 
\subsection{$|\Delta_S|$ and reality
 \label{sec:BRSTDsol}}

\begin{figure}[htbp]
 \centering
 \includegraphics[height=6.5cm]{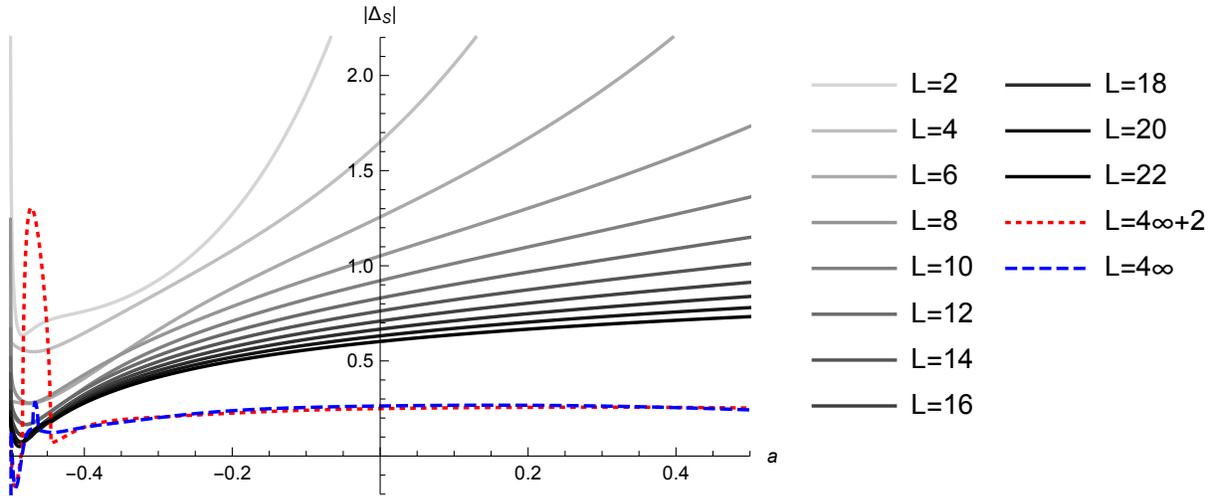}
 \caption{
 Plots of $|\Delta_S|$ (\ref{eq:absDelta_S}) 
 for the  ``double brane" solution $\Phi_a^{\rm D}$
 at the truncation level $L$.
 The horizontal axis denotes the value of the parameter $a$.
 \label{fig:delta_s_dsol}
 }
\end{figure}

Fig.~\ref{fig:delta_s_dsol} shows plots of $|\Delta_S|$
(\ref{eq:absDelta_S}) for the ``double brane" solution $\Phi_a^{\rm D}$
at the truncation level $L$.  With increasing level, it approaches zero
for $a\ge -1/2$.  This behavior is consistent with the equation of
motion (\ref{eq:EOMQp}) for $\Phi_a^{\rm D}$.  The extrapolation values
are close to $0.25$ for $a>-1/2$ except around $a=-1/2$, where they are
unstable.  We could expect that the above does not contradict that
$|\Delta_S|\to 0$ ($L\to \infty$) for $a\ge -1/2$.
 
\begin{figure}[htbp]
 \centering
 \includegraphics[height=6.5cm]{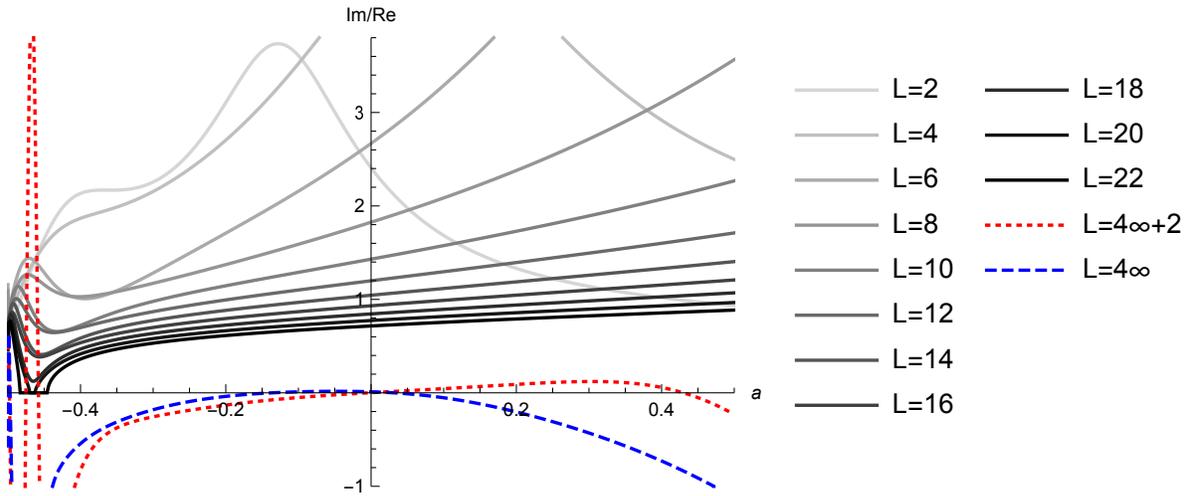}
 \caption{
 Plots of $\text{Im}/\text{Re}$ (\ref{eq:Im/RePhi_a}) 
 for the  ``double brane" solution $\Phi_a^{\rm D}$ at 
 the truncation level $L$. The horizontal axis denotes 
 the value of the parameter $a$. 
 \label{fig:ior_dsol}
 }
\end{figure}

\begin{figure}[htbp]
 \centering
 \includegraphics[height=6.5cm]{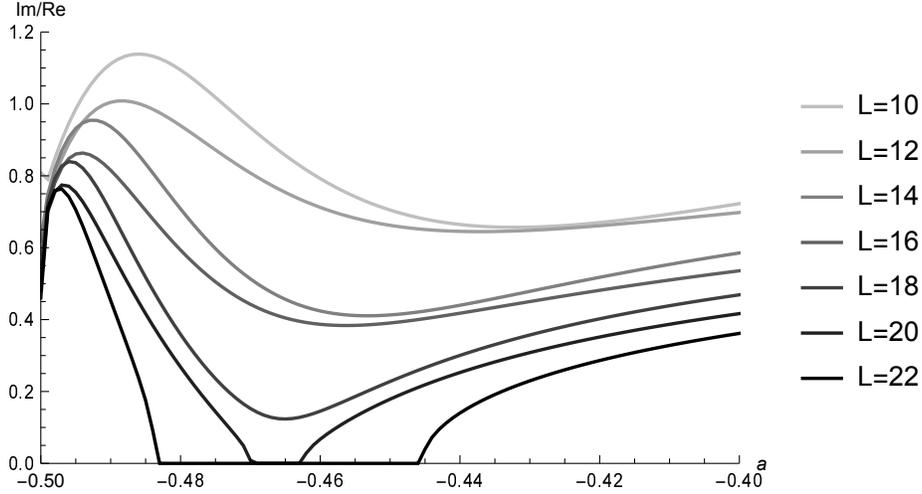}
 \caption{
 Plots of $\text{Im}/\text{Re}$ (\ref{eq:Im/RePhi_a}) for
 the  ``double brane" 
 solution $\Phi_a^{\rm D}$ such as $a\gtrsim -1/2$
 at the truncation level $L=10,12,\cdots,22$.
 The horizontal axis denotes the value of the parameter $a$.
 The vertical axis stands at $a=-1/2$.
 \label{fig:ior_dsol10}}
\end{figure}

Figs.~\ref{fig:ior_dsol} and \ref{fig:ior_dsol10} show plots of
$\text{Im}/\text{Re}$ (\ref{eq:Im/RePhi_a}) for the ``double brane"
solution $\Phi_a^{\rm D}$ at the truncation level $L$.  With increasing
level, it approaches zero for $a\ge -1/2$.  This behavior is consistent
with the reality condition of the string field $\Phi_a^{\rm D}$.  In
particular, from Fig.~\ref{fig:ior_dsol10}, we have found that
$\text{Im}/\text{Re}=0$ for $-0.469\le a\le -0.463$ at $L=20$ and for
$-0.483\le a\le -0.446$ at $L=22$ (although there is no solution at
$a=-0.468$ in both cases as mentioned).  The region of $a$ at $L=22$,
where $\Phi_a^{\rm D}$ is real, is larger than that at $L=20$.  We can
expect that it becomes larger at higher truncation level because
$\Phi_a^{\rm D}$ at the level $L+2$ is real once it becomes real at the
level $L$ from our method to construct the solutions.  We note that the
regions of $a$ such that $\Phi_a^{\rm D}$ is real correspond to those
where the value of ${\rm Re}\,E_0[\Phi_a^{\rm D}]$ shows irregular
behavior in Fig.~\ref{fig:re_e0_dsol10}.

In Fig.~\ref{fig:ior_dsol}, the extrapolation values become negative for
some region of $a$ although $\text{Im}/\text{Re}$ should be nonnegative by
its definition (\ref{eq:Im/RePhi_a}).  We might interpret that
$\Phi_a^{\rm D}$ becomes real at finite level in such a region.

From the above observations, we expect that the ``double brane" solution
$\Phi_a^{\rm D}$ for $a\ge -1/2$ satisfies the reality condition in the
large level limit $L\to \infty$.

 
%%%%%%%%%%%%%%%
\section{
 ``Ghost brane" solution
 \label{sec:Gsol}
 }
 
Our strategy to construct solutions corresponding to ``ghost brane" is
the same as that in \S \ref{sec:Dsol}.
 
In the theory around the perturbative vacuum, which is the case $a=0$ in
(\ref{eq:S_aPhi}), a ``ghost brane" solution $\Phi^{\rm G}_{a=0}$ is
obtained from one of complex solutions at the truncation level $4$.
Taking a solution $\Phi^{\rm G}_{a=0}$ at level $L$ as an initial string
field, we can obtain $\Phi^{\rm G}_{a=0}$ at level $L+2$ by Newton's
method.  Such a solution $\Phi^{\rm G}_{a=0}$ coincides with the ``ghost
brane" solution in \cite{Kudrna:2018mxa}.
 
Firstly, at the truncation level 4, we constructed $\Phi^{\rm G}_{a}$
($-1/2\le a\le 1/2$) from $\Phi^{\rm G}_{a=0}$.  Then, for a fixed value
of $a$, which is one of $a=-0.5,-0.499,-0.498,\cdots,0.499,0.5$, we
constructed higher level solutions from $\Phi^{\rm G}_a$ at level 4 up
to $22$.  As a result, we have obtained numerical solutions $\Phi^{\rm
G}_a$, except for $\Phi^{\rm G}_{a=-0.499}$ at level $22$.  In the case
$a=-0.499$, Newton's method for constructing a solution at level 22 did
not converge.

\subsection{Energy
\label{sec:EGsol}
}

\begin{figure}[htbp]
 \centering
 \includegraphics[height=6.5cm]{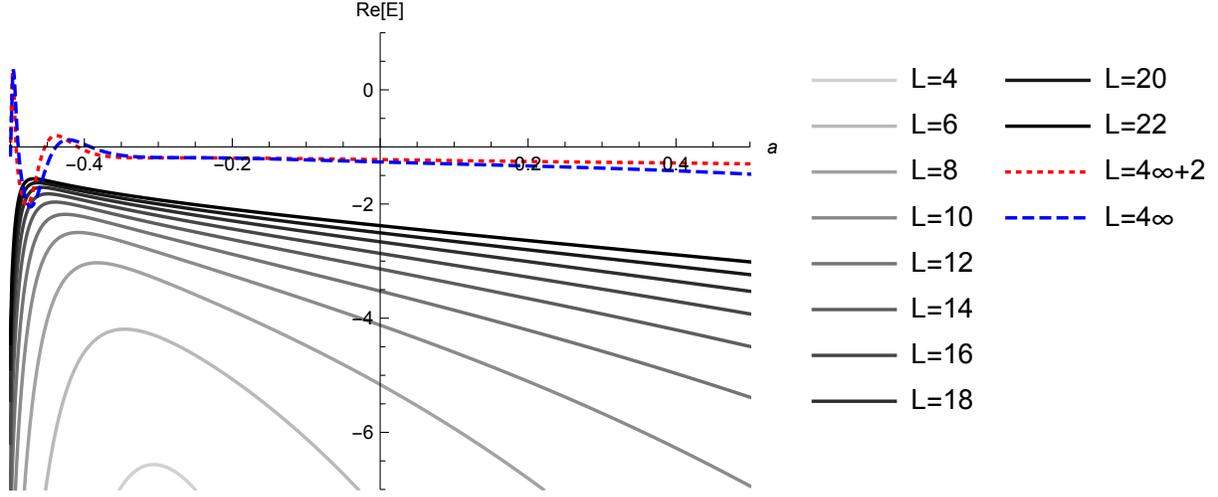}
 \caption{
 Plots of the real part of the energy $E$ (\ref{eq:EPhia}) for the
 ``ghost brane" solution $\Phi_a^{\rm G}$ at the truncation level $L$.
 The dotted and dashed lines are extrapolations to $L=4k+2$ ($k\to
 \infty$) and $L=4k$ ($k\to \infty$), respectively.  The horizontal axis
 denotes the value of the parameter $a$ at ${\rm Re}\,E=-1$.
 \label{fig:re_e_gsol} }
\end{figure}

\begin{figure}[htbp]
 \centering 
 \includegraphics[height=6.5cm]{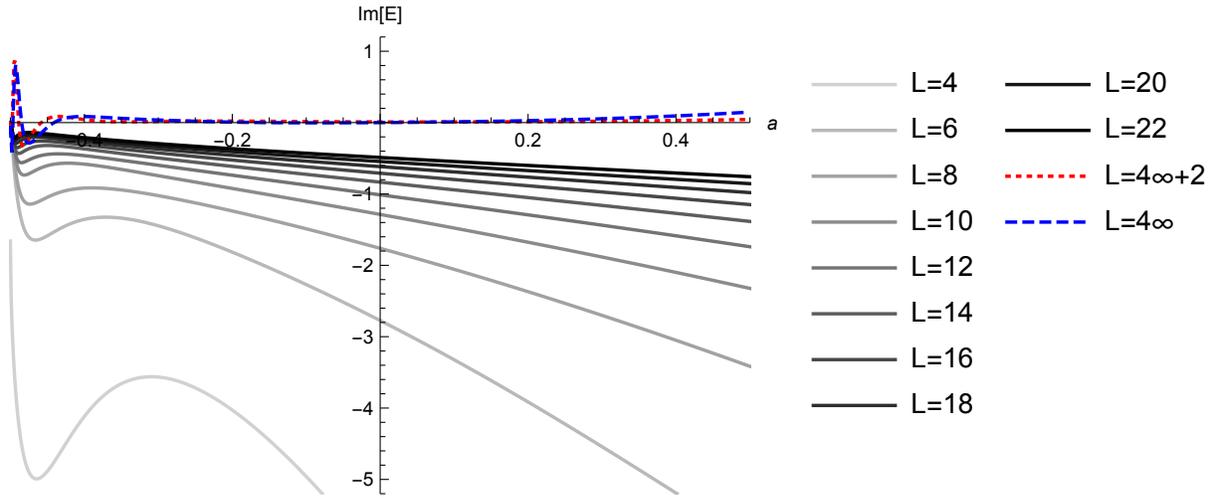}
 \caption{
 Plots of the imaginary part of the energy $E$ (\ref{eq:EPhia}) for the
 ``ghost brane" solution $\Phi_a^{\rm G}$ at the truncation level $L$.
 The dotted and dashed lines are extrapolations to $L=4k+2$ ($k\to
 \infty$) and $L=4k$ ($k\to \infty$), respectively.  The horizontal axis
 denotes the value of the parameter $a$.  \label{fig:im_e_gsol} }
 \end{figure}

Figs.~\ref{fig:re_e_gsol} and \ref{fig:im_e_gsol} show plots of the
energy $E$ (\ref{eq:EPhia}) for the ``ghost brane" solution $\Phi_a^{\rm
G}$ at the truncation level $L$.  As in Fig.~\ref{fig:re_e_gsol}, the
real part ${\rm Re}\,E[\Phi_a^{\rm G}]$ increases with increasing level.
The extrapolations are close to $-1$ for
$a>-1/2$.  There is a maximum near $a=-1/2$ for each extrapolation and
the maximum value is greater than zero.

As in Fig.~\ref{fig:im_e_gsol}, the imaginary part ${\rm Im}\,E[\Phi_a^{\rm G}]$ 
approaches zero with increasing level.  In particular, extrapolation values are
close to zero for $a>-1/2$ although they become unstable around
$a=-1/2$.  We note that ${\rm Im}\,E[\Phi_a^{\rm G}]=0$ at $a=-1/2$ for
the level $L\ge 6$ as in Table~\ref{tab:a=m1o2G} in appendix~\ref{sec:App}.
 
From the above, we might expect that the energy for the ``ghost brane"
solution behaves as
\begin{align}
&E[\Phi_a^{\rm G}]\to 
\begin{cases}
E_{-1} &(a>-1/2)\\
E_{-1}^{\prime} &(a=-1/2)
\end{cases}
&&(L\to \infty),
\label{eq:EaPsiGE2E2p}
\end{align}
where $E_{-1}$ and $E_{-1}^{\prime}$ seem to be real constants such as
$-2<E_{-1}<E_{-1}^{\prime}$, in a similar way to (\ref{eq:EaPsiDE2E2p})
for $\Phi_a^{\rm D}$, but it is more obscure in this case.  If we focus
on the extrapolation values, it seems that $E_{-1}\sim -1.25$ and
$0<E_{-1}^{\prime}<1$, although $\Phi_a^{\rm G}$ could represent ghost
brane literally if $E_{-1}=-1$ and $E_{-1}^{\prime}=0$.  If
$E_{-1}^{\prime}-E_{-1}=1$, $\Phi_a^{\rm G}$ for $a\ge -1/2$ is
consistent with the $a$-dependence of the TT solution.

\subsection{Gauge invariant observable
\label{sec:E0Gsol}}
 
\begin{figure}[htbp]
 \centering 
 \includegraphics[height=6.5cm]{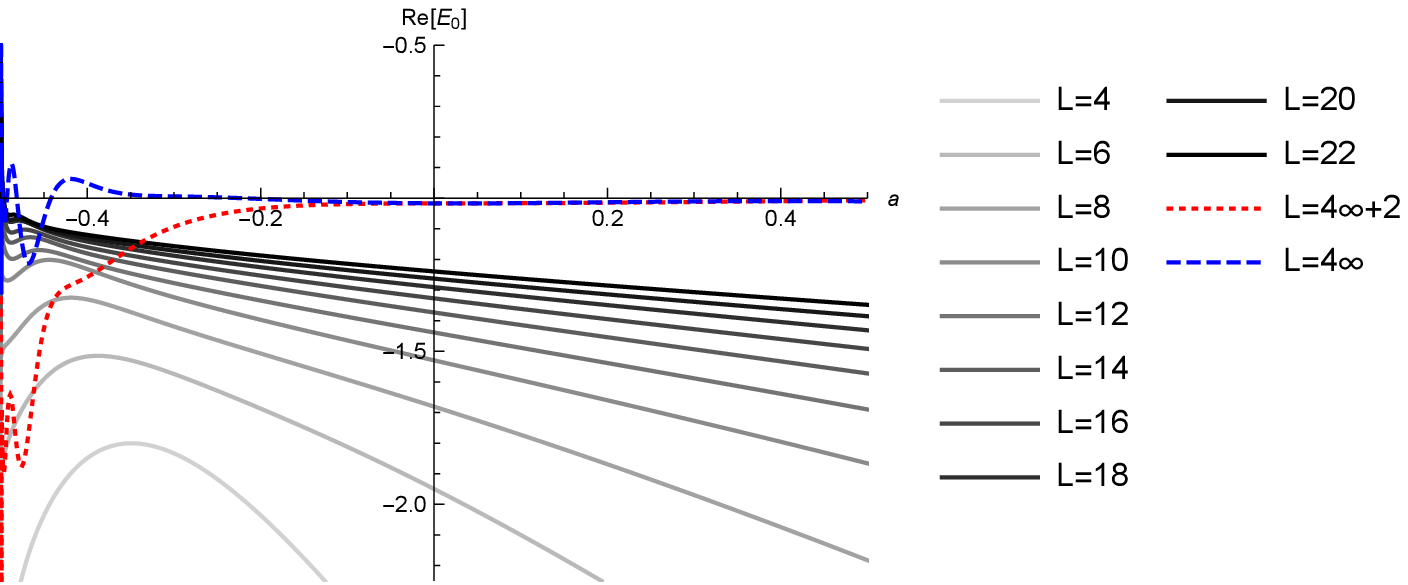} 
 \caption{
 Plots of the real part of the gauge invariant observable $E_0$
 (\ref{eq:E0Phia}) for the ``ghost brane" solution $\Phi_a^{\rm G}$ at
 the truncation level $L$.  The dotted and dashed lines are
 extrapolations to $L=4k+2$ ($k\to \infty$) and $L=4k$ ($k\to \infty$),
 respectively.  The horizontal axis denotes the value of the parameter
 $a$ at ${\rm Re}\,E_0=-1$.  \label{fig:re_e0_gsol} 
}
\end{figure}

\begin{figure}[htbp]
  \centering
  \includegraphics[height=6.5cm]{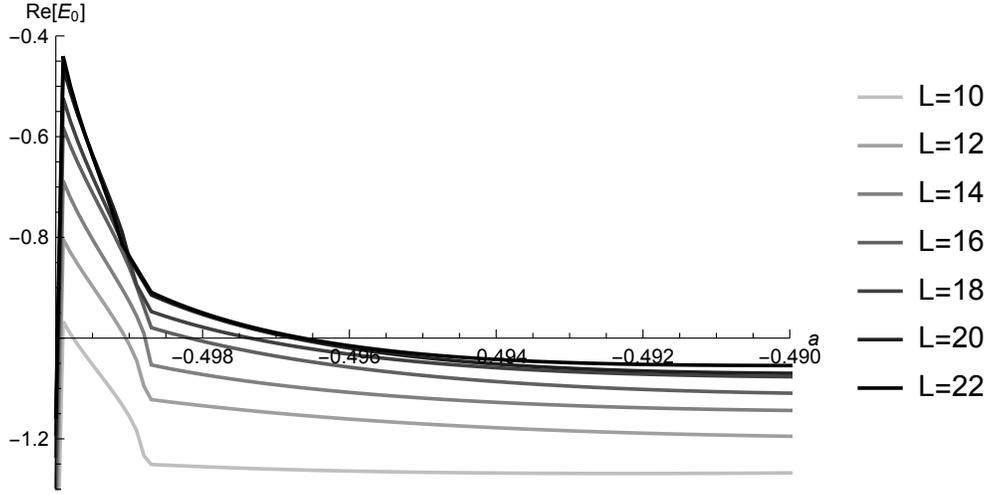}
  \caption{
 Plots of the real part of the gauge invariant observable $E_0$
 (\ref{eq:E0Phia}) for the ``ghost brane" solution $\Phi_a^{\rm G}$ near
 $a= -1/2$ at the truncation level $L=10,12,\cdots,22$.  The horizontal
 axis denotes the value of the parameter $a$ at ${\rm Re}\,E_0=-1$.  The
 vertical axis stands at $a=-1/2$.  \label{fig:re_e0_gsol10} }
\end{figure}

\begin{figure}[htbp]
 \centering
 \includegraphics[height=6.5cm]{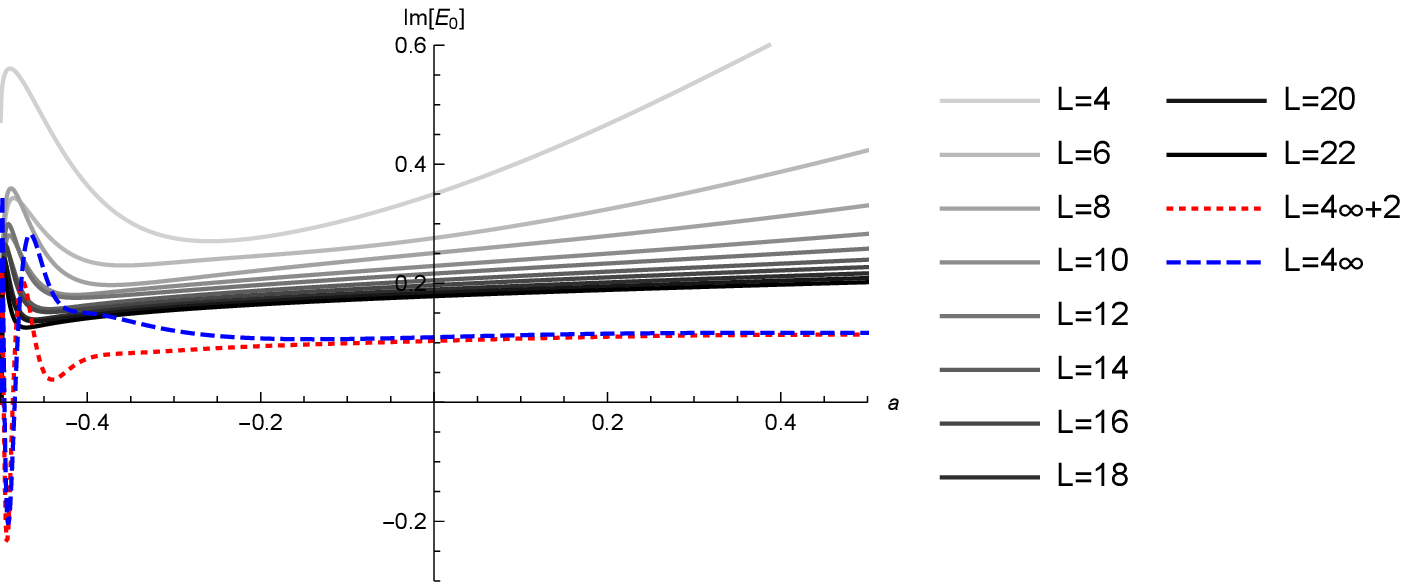}
 \caption{
 Plots of the imaginary part of the gauge invariant observable $E_0$
 (\ref{eq:E0Phia}) for the ``ghost brane" solution $\Phi_a^{\rm G}$ at
 the truncation level $L$.  The dotted and dashed lines are
 extrapolations to $L=4k+2$ ($k\to \infty$) and $L=4k$ ($k\to \infty$),
 respectively.  The horizontal axis denotes the value of the parameter
 $a$.  \label{fig:im_e0_gsol} }
\end{figure}

Figs.~\ref{fig:re_e0_gsol}, \ref{fig:re_e0_gsol10}, and
\ref{fig:im_e0_gsol} show plots of the gauge invariant observable $E_0$
(\ref{eq:E0Phia}) for the ``ghost brane" solution $\Phi_a^{\rm G}$ at
the truncation level $L$.  As in Fig.~\ref{fig:re_e0_gsol}, the real
part ${\rm Re}\,E_0[\Phi_a^{\rm G}]$ approaches $-1$ for $a>-1/2$.  As
in Fig.~\ref{fig:re_e0_gsol10}, there is a maximum near $a=-1/2$ for
each level $L$.  We have found that the maximum values are greater than
$-1$ for $L\ge 10$ although there is a large error in extrapolation
values of ${\rm Re}\,E_0[\Phi_a^{\rm G}]$ near $a=-1/2$ in
Fig.~\ref{fig:re_e0_gsol}.\footnote{ We performed computations at
$a=-0.5,-0.4999,-0.4998,\cdots,-0.4901,-0.49$ in order to show details
near $a=-0.5$ in Fig.~\ref{fig:re_e0_gsol10} (and
Fig.~\ref{fig:re_e0_gsol}).  In addition to $a=-0.499$ for $L=22$, we
did not obtain solutions at $a=-0.4989$ for $L=20,22$ and at $a=-0.4988$
for $L=16,18,20,22$.  }

As in Fig.~\ref{fig:im_e0_gsol}, the imaginary part ${\rm
Im}\,E_0[\Phi_a^{\rm G}]$ decreases with increasing level.  Its
extrapolation values are about $0.1$ for $a>-1/2$ and they become
unstable around $a=-1/2$.  At $a=-1/2$, ${\rm Im}\,E_0[\Phi_a^{\rm
G}]=0$ for the level $L\ge 6$ as in Table \ref{tab:a=m1o2G} in appendix~\ref{sec:App}.
 
From the above, in a similar way to (\ref{eq:EaPsiGE2E2p}), we could
expect that the gauge invariant observable (\ref{eq:E0Phia}) behaves as
\begin{align}
E_0[\Phi_a^{\rm G}]\to 
\begin{cases}
\tilde E_{-1} &(a>-1/2)\\
\tilde E_{-1}^{\prime} &(a=-1/2)
\end{cases}
&&(L\to \infty),
\label{eq:E0PhiaGE2E2p}
\end{align}
where $\tilde E_{-1}$ and $\tilde E_{-1}^{\prime}$ 
seem to be real constants such as $-2<\tilde E_{-1}<\tilde E_{-1}^{\prime}$.

\subsection{$|\Delta_S|$ and reality
 \label{sec:BRSTGsol}}
 
\begin{figure}[htbp]
 \centering
 \includegraphics[height=6.5cm]{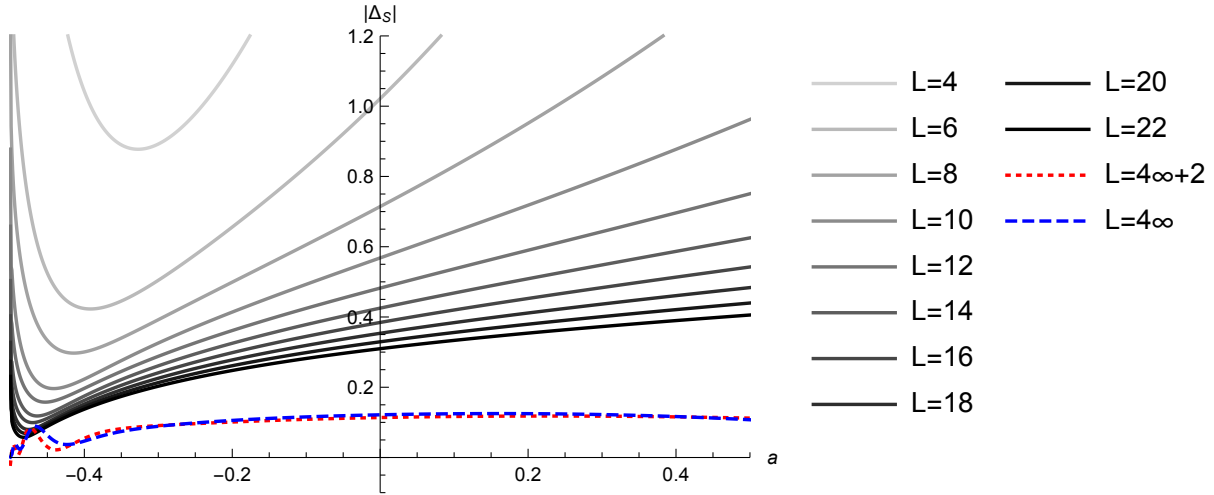}
 \caption{
 Plots of $|\Delta_S|$ (\ref{eq:absDelta_S}) 
 for the  ``ghost brane" solution $\Phi_a^{\rm G}$
 at the truncation level $L$.
 The horizontal axis denotes the value of the parameter $a$.
 \label{fig:delta_s_gsol}
 }
 \end{figure}

Fig.~\ref{fig:delta_s_gsol} shows plots of $|\Delta_S|$
(\ref{eq:absDelta_S}) for the ``ghost brane" solution $\Phi_a^{\rm G}$
at the truncation level $L$.  It decreases with increasing level.  The
extrapolation values are about $0.1$ for $a\gtrsim -1/2$ and they are
close to zero near $a=-1/2$.  It is consistent with the equation of
motion (\ref{eq:EOMQp}) for $\Phi_a^{\rm G}$.
 
\begin{figure}[htbp]
 \centering
 \includegraphics[height=6.5cm]{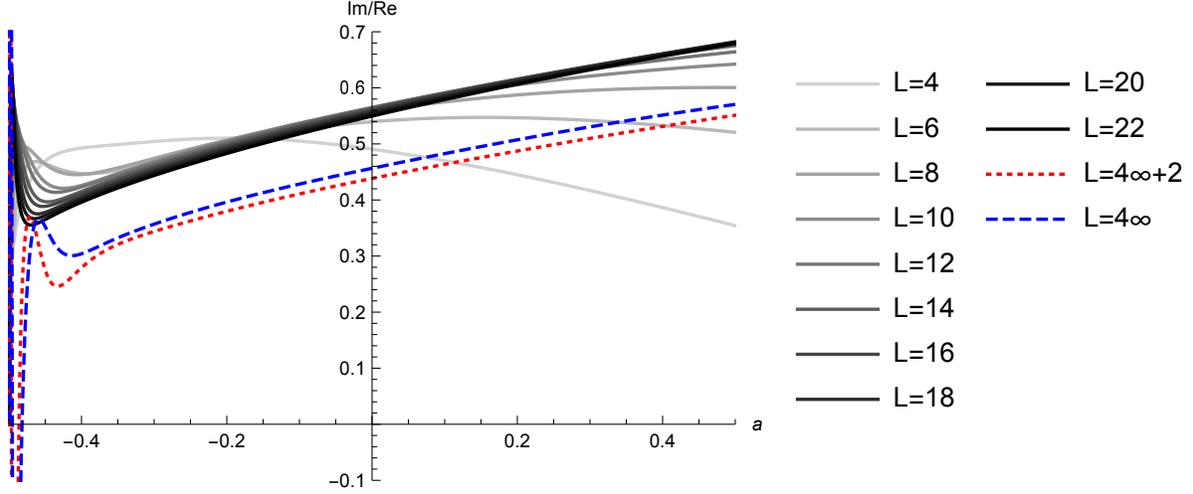}
 \caption{
 Plots of $\text{Im}/\text{Re}$ (\ref{eq:Im/RePhi_a}) 
 for the  ``ghost brane" solution $\Phi_a^{\rm G}$
 at the truncation level $L$.
 The horizontal axis denotes the value of the parameter $a$.
 \label{fig:ior_gsol}
 }
\end{figure}

Fig.~\ref{fig:ior_gsol} shows plots of $\text{Im}/\text{Re}$
(\ref{eq:Im/RePhi_a}) for the ``ghost brane" solution $\Phi_a^{\rm G}$
at the truncation level $L$.  For $a\gtrsim -0.4$, $\text{Im}/\text{Re}$
at level $L$ up to $22$ and its extrapolation values are greater than
$0.2$.  On the other hand, we found that
$\text{Im}/\text{Re}[\Phi_a^{\rm G}]=0$ at $a=-0.5,-0.499$ for $L\ge 6$,
although there is no solution at $a=-0.499$ for $L=22$.
 
In any case, it seems that $\Phi_a^{\rm G}$ does not satisfy the reality
condition for $a\gtrsim -0.4$ even at the limit $L\to \infty$ from the
numerical behavior of $\text{Im}/\text{Re}[\Phi_a^{\rm G}]$ in
Fig.~\ref{fig:ior_gsol} and comparison with that for $\Phi_a^{\rm D}$ in
Fig.~\ref{fig:ior_dsol}.  It is consistent with the interpretation of
the ``ghost brane" solution in \cite{Kudrna:2018mxa}, which corresponds
to the case $a=0$.

\section{
 Concluding remarks
 \label{sec:remarks}
 }
 
In this paper, we have numerically constructed the ``double brane"
solution $\Phi_a^{\rm D}$ and the ``ghost brane" solution $\Phi_a^{\rm
G}$ in the theory around the TT solution with a real parameter $a$ by using 
the level truncation method.  In particular, in the case $a=0$, they
coincide with the ``double brane" and ``ghost brane" solutions found by
Kudrna and Schnabl.  In this sense, our solutions are generalization of
theirs.  We have evaluated the energy $E$ and gauge invariant observable
$E_0$ for the obtained solutions and calculated $|\Delta_S|$ and
$\text{Im}/\text{Re}$ as a consistency check.
 
From our results, in the large level limit, it seems that $E[\Phi_a^{\rm
D}]\to E_2$ for $a>-1/2$ and $\to E_2^{\prime}$ for $a=-1/2$, where the
real constants, $E_2$ and $E_2^{\prime}$, satisfy $1<E_2<E_2^{\prime}$.
If $E_2=2=E_2^{\prime}-1$, $\Phi_a^{\rm D}$ could be interpreted as
double brane solution for $a\ge -1/2$ literally, but we observed
$E_2\sim 1.5$ and $2<E_2^{\prime}<3$ from our numerical result.
However, the solution $\Phi_a^{\rm D}$ for $a\ge -1/2$ is consistent
with the interpretation that the TT solution represents the tachyon
vacuum at $a=-1/2$ and is pure gauge for $a>-1/2$ if
$E_2^{\prime}-E_2=1$, which may hold.  The value of $E_0[\Phi_a^{\rm
D}]$ behaves in a similar way to $E[\Phi_a^{\rm D}]$ numerically.  In
order to establish the relation $E_2^{\prime}-E_2=1$, detailed
computations around $a=-1/2$ for higher levels will be necessary.  As
for $\Phi_a^{\rm G}$, numerical behavior of the energy and gauge
invariant observable is similar to that of $\Phi_a^{\rm D}$, but it is
more ambiguous.
  
We observed that both $|\Delta_S[\Phi_a^{\rm D}]|$ and
$|\Delta_S[\Phi_a^{\rm G}]|$ approach zero with increasing level, which
is consistent with the projected equation of motion:
$b_0c_0(Q^{\prime}\Phi+\Phi\ast\Phi)=0$ up to the lowest level.  We
should check coefficients of higher level states in order to confirm the
BRST invariance of solutions in Siegel gauge.  It remains as a future
work.
  
The numerical solutions $\Phi_a^{\rm D}$ and $\Phi_a^{\rm G}$ are
constructed from complex solutions at level 2 and 4, respectively.
Namely, they do not satisfy the reality condition of string field and
therefore $E$ and $E_0$ for them have imaginary part in general.
However, we observed that $\text{Im}/\text{Re}[\Phi_a^{\rm D}]\to 0$
with increasing level for $a\ge -1/2$.  In particular, we found a region
of $a$ where $\text{Im}/\text{Re}[\Phi_a^{\rm D}]=0$ at the level $20$
and $22$.  In this sense, $\Phi_a^{\rm D}$ is expected to be real at the
large level limit for $a\ge -1/2$.  As for $\Phi_a^{\rm G}$, it seems
that $\text{Im}/\text{Re}[\Phi_a^{\rm G}]$ does not approach zero with
increasing level for $a\gtrsim -0.4$, although we found that at $a=-0.5$
and $a=-0.499$, $\text{Im}/\text{Re}[\Phi_a^{\rm G}]=0$ for the level $L\ge 6$.
 
We have constructed numerical solutions by Newton's method, where we
have to choose appropriate initial configurations for the iterative
algorithm.  In this paper, we have adopted a choice explained in \S
\ref{sec:Dsol}:~ $1$.~At the lowest truncation level, which is two for
$\Phi_a^{\rm D}$ and four for $\Phi_a^{\rm G}$, we have constructed
solutions for various values of $a$ from $\Phi_{a=0}^{\rm D}$ and
$\Phi_{a=0}^{\rm G}$ by varying the value of $a$ little by little.
~$2$.~For each value of $a$, higher level solutions have been
constructed level by level.~ However, we can make another choice of
initial configurations:~ $1^{\prime}$.~For $a=0$, we construct higher
level solutions level by level.  ~$2^{\prime}$.~At each level, 
solutions for $a\ne 0$ are constructed from that for $a=0$ by varying the
value of $a$ little by little.~ We have observed that there is a
possibility of obtaining different solutions by
$1^{\prime}$-$2^{\prime}$ from those by $1$-$2$, where solutions become
real near $a=-1/2$.  Further detailed investigation is a future
problem although the procedure $1^{\prime}$-$2^{\prime}$ takes more time
for calculation.
 
In this paper, we have adopted a method of extrapolations, which is
explained in \S \ref{sec:EDsol}.  There are two extrapolation values:
one is for the truncation level $L=4\infty+2$ and the other is for
$L=4\infty$, which are plotted in Figures by dotted and dashed lines.
The difference of them would correspond to an error of the
extrapolation.  We should justify the extrapolation method or adopt
other methods in order to refine the numerical results.
 
We have performed calculations in Siegel gauge for simplicity.  It
is desirable to evaluate gauge invariants in other gauges for ``double
brane" and ``ghost brane" solutions to understand them further.  We are
preparing computations of them in Asano-Kato gauge \cite{Asano:2006hk}
as in \cite{Kishimoto:2009cz, Kishimoto:2009hb}.\footnote{ Some comments
can be found on calculations in Schnabl gauge in \cite{Arroyo:2019liw}.
}
%%%%%%%%%%%%%%%%%%%%%%%%%%%%%%%%%%%%%%%
\section*{Acknowledgments}
This work was supported in part by JSPS KAKENHI Grant Numbers
JP20K03933, JP20K03972.  The numerical calculations were partly carried
out on sushiki and XC40 at YITP in Kyoto University.
%%%%%%%%%%%%%%%%%%%%%%%%%%%%%%%%%%%%%%%

\appendix
%%%%%%%%%%%%%%%%%%%%%%%%%%%%%%%%%%%%%%%
\section{Some numerical data for solutions
\label{sec:App}
}

\subsection{Numerical data for solutions at $a=-1/2$
}
We list explicit numerical values of solutions 
in the theory around the TT solution at $a=-1/2$
in Tables \ref{tab:a=m1o2D} and \ref{tab:a=m1o2G}.
They can be compared with those in \cite{Kudrna:2018mxa},
which correspond to the case $a=0$.\footnote{
Our definition of $\text{Im}/\text{Re}$ seems to be different from that in \cite{Kudrna:2018mxa}.
We suspect that a choice of a basis (or its normalization) is  different.
}

The data in Table  \ref{tab:a=m1o2D} are
$E$ (\ref{eq:EPhia}), $E_0$ (\ref{eq:E0Phia}), $|\Delta_S|$ (\ref{eq:absDelta_S}), 
and $\text{Im}/\text{Re}$ (\ref{eq:Im/RePhi_a}) for  the ``double brane" solution $\Phi^{\rm D}_{a=-1/2}$ 
and they correspond to points  at $a=-1/2$ 
in Figs.~\ref{fig:re_e_dsol}, \ref{fig:re_e_dsol10}, \ref{fig:im_e_dsol},
\ref{fig:re_e0_dsol}, \ref{fig:re_e0_dsol10}, \ref{fig:im_e0_dsol}, \ref{fig:delta_s_dsol}, 
\ref{fig:ior_dsol}, and \ref{fig:ior_dsol10}.
The data in Table  \ref{tab:a=m1o2G} are
$E$ (\ref{eq:EPhia}), $E_0$ (\ref{eq:E0Phia}), $|\Delta_S|$ (\ref{eq:absDelta_S}), 
and $\text{Im}/\text{Re}$ (\ref{eq:Im/RePhi_a}) for  the ``ghost brane" solution $\Phi^{\rm G}_{a=-1/2}$ 
and they correspond to points  at $a=-1/2$ 
in Figs.~\ref{fig:re_e_gsol}, \ref{fig:im_e_gsol},
\ref{fig:re_e0_gsol}, \ref{fig:re_e0_gsol10}, \ref{fig:im_e0_gsol}, \ref{fig:delta_s_gsol}, and
\ref{fig:ior_gsol}.
\begin{table}[htbp]
        \caption{
        $E$ (\ref{eq:EPhia}), $E_0$ (\ref{eq:E0Phia}), $|\Delta_S|$ (\ref{eq:absDelta_S}),
        and  $\text{Im}/\text{Re}$ (\ref{eq:Im/RePhi_a}) for  the ``double brane" solution $\Phi^{\rm D}_{a=-1/2}$ at the truncation level $L$.
    \label{tab:a=m1o2D}}
    \begin{center}
%  %   \renewcommand{\arraystretch}{1.3}
\begin{tabular}{|r|ll|ll|}
\hline
$L$&$E$&$E_0$&$|\Delta_S|$&$\text{Im}/\text{Re}$\\
\hline
2&$-0.738519 + 6.42573 i$
&$2.5906 - 0.611421 i$
&$2.27577$
&$0.414127$
\\
%%\hline
4&$-0.520622 - 3.02656 i$
&$2.7175 - 0.100579 i$
&$0.624772$
&$0.664631$
\\
%%\hline
6&$-0.810759 - 1.90909 i$
&$2.62035 - 0.100411 i$
&$0.781281$
&$1.15795$
\\
%%\hline
8&$1.2739 - 1.27992 i$
&$2.59001 - 0.122603 i$
&$1.24412$
&$1.09292$
\\
%%\hline
10&$0.947834 - 1.30151 i$
&$2.38848 + 0.0741147 i$
&$0.667769$
&$0.80842$
\\
%%\hline
12&$1.55586 - 1.07501 i$
&$2.35887 + 0.0660807 i$
&$0.517581$
&$0.57895$
\\
%%\hline
14&$1.51726 - 0.904625 i$
&$2.2996 + 0.0707516 i$
&$0.431135$
&$0.617605$
\\
%%\hline
16&$1.78862 - 0.768909 i$
&$2.25904 + 0.0906131 i$
&$0.339793$
&$0.501388$
\\
%%\hline
18&$1.7433 - 0.660931 i$
&$2.22563 + 0.0962804 i$
&$0.283798$
&$0.531292$
\\
%%\hline
20&$1.89055 - 0.579279 i$
&$2.21258 + 0.0902456 i$
&$0.232422$
&$0.460356$
\\
%%\hline
22&$1.85117 - 0.508455 i$
&$2.19168 + 0.0854377 i$
&$0.195764$
&$0.481028$
\\
\hline
\end{tabular}
% %      \renewcommand{\arraystretch}{1}
    \end{center}
\end{table}
\begin{table}[htbp]
\caption{
$E$ (\ref{eq:EPhia}), $E_0$ (\ref{eq:E0Phia}), $|\Delta_S|$ (\ref{eq:absDelta_S}), and
 $\text{Im}/\text{Re}$ (\ref{eq:Im/RePhi_a}) for  the ``ghost brane" solution $\Phi^{\rm G}_{a=-1/2}$ 
 at the truncation level $L$. 
 $\Phi^{\rm G}_{a=-1/2}$ satisfies the reality condition for $L\ge 6$.
    \label{tab:a=m1o2G}}
    \begin{center}
%  %   \renewcommand{\arraystretch}{1.3}
\begin{tabular}{|r|ll|ll|}
\hline
$L$&$E$&$E_0$&$|\Delta_S|$&$\text{Im}/\text{Re}$\\
\hline
4&$-58.3862 - 1.66004 i$
&$-2.95681 + 0.472707 i$
&$4.66418$
&$0.189811$
\\
%%\hline
6&$-24.413$
&$-2.19698$
&$1.91163$
&$0$
\\
%%\hline
8&$-15.6787$
&$-1.81939$
&$1.29319$
&$0$
\\
%%\hline
10&$-11.1078$
&$-1.58933$
&$0.87768$
&$0$
\\
%%\hline
12&$-8.76291$
&$-1.46405$
&$0.65827$
&$0$
\\
%%\hline
14&$-7.20365$
&$-1.37387$
&$0.503632$
&$0$
\\
%%\hline
16&$-6.19244$
&$-1.29508$
&$0.40301$
&$0$
\\
%%\hline
18&$-5.43759$
&$-1.23428$
&$0.327863$
&$0$
\\
%%\hline
20&$-4.88756$
&$-1.19106$
&$0.273432$
&$0$
\\
%%\hline
22&$-4.44884$
&$-1.15705$
&$0.230936$
&$0$
\\
\hline
\end{tabular}
% %      \renewcommand{\arraystretch}{1}
    \end{center}
\end{table}

\subsection{Coefficients of lowest level states for the solutions}

\begin{figure}[htbp]
 \centering
 \includegraphics[height=6.5cm]{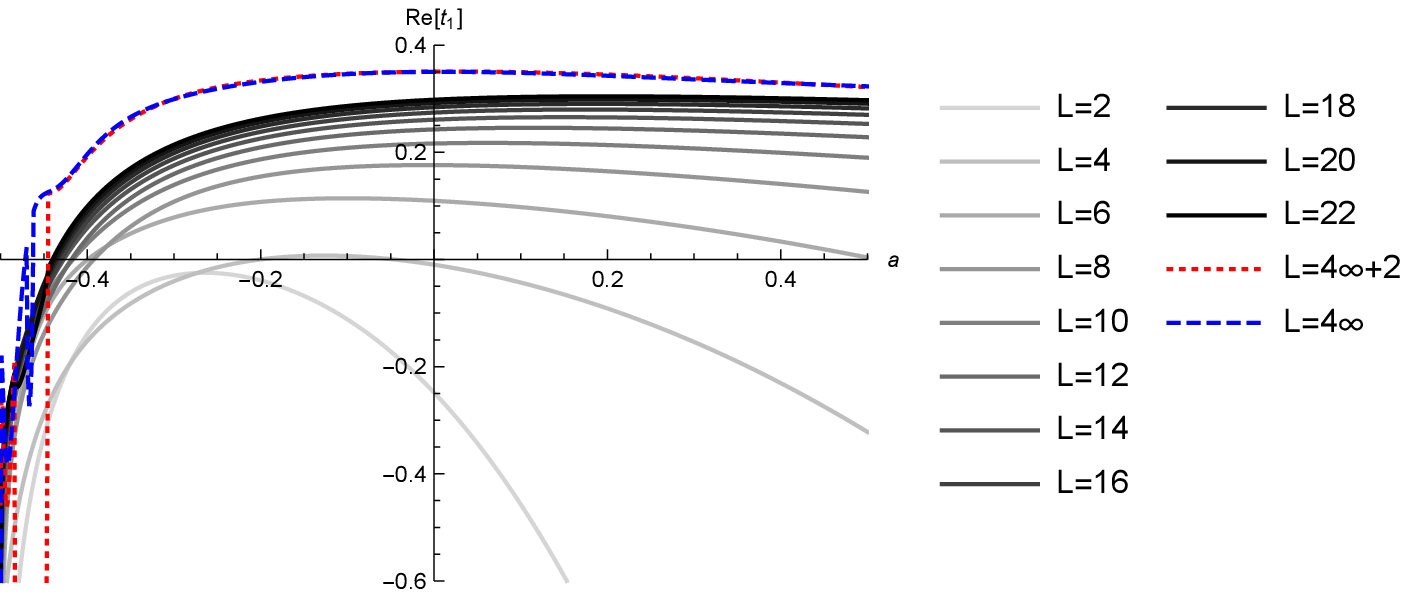}
 \caption{
 Plots of the real part of the coefficient of the lowest level state $c_1|0\rangle$
 for the ``double brane" solution $\Phi^{\rm D}_a$ at the truncation level $L$.
 The dotted and dashed lines are extrapolations to $L=4k+2$ ($k\to \infty$) and $L=4k$ ($k\to \infty$),
 respectively.
 The horizontal axis denotes the values of the parameter $a$.
  \label{fig:re_t1_dsol}
 }
 \end{figure}

\begin{figure}[htbp]
 \centering
 \includegraphics[height=6.5cm]{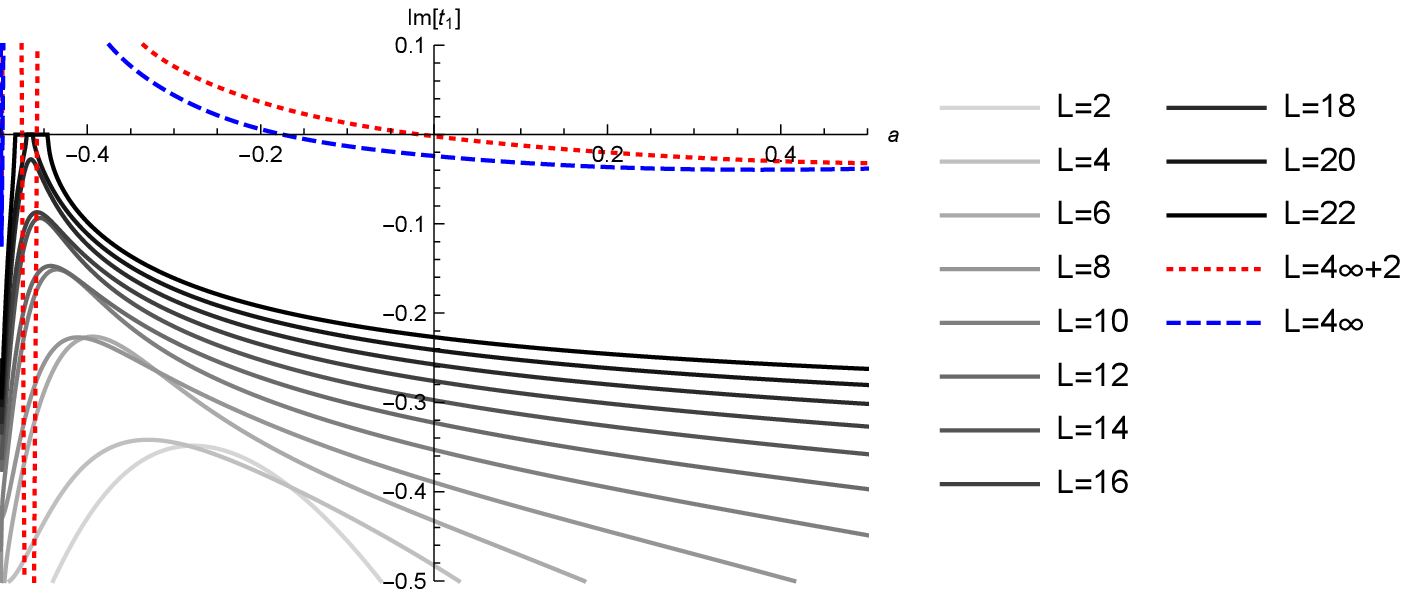}
 \caption{
 Plots of the imaginary part of the coefficient of the lowest level state $c_1|0\rangle$
 for the ``double brane" solution $\Phi^{\rm D}_a$ at the truncation level $L$.
 The dotted and dashed lines are extrapolations to $L=4k+2$ ($k\to \infty$) and $L=4k$ ($k\to \infty$),
 respectively.
 The horizontal axis denotes the values of the parameter $a$.
  \label{fig:im_t1_dsol}
 }
 \end{figure}

\begin{figure}[htbp]
 \centering
 \includegraphics[height=6.5cm]{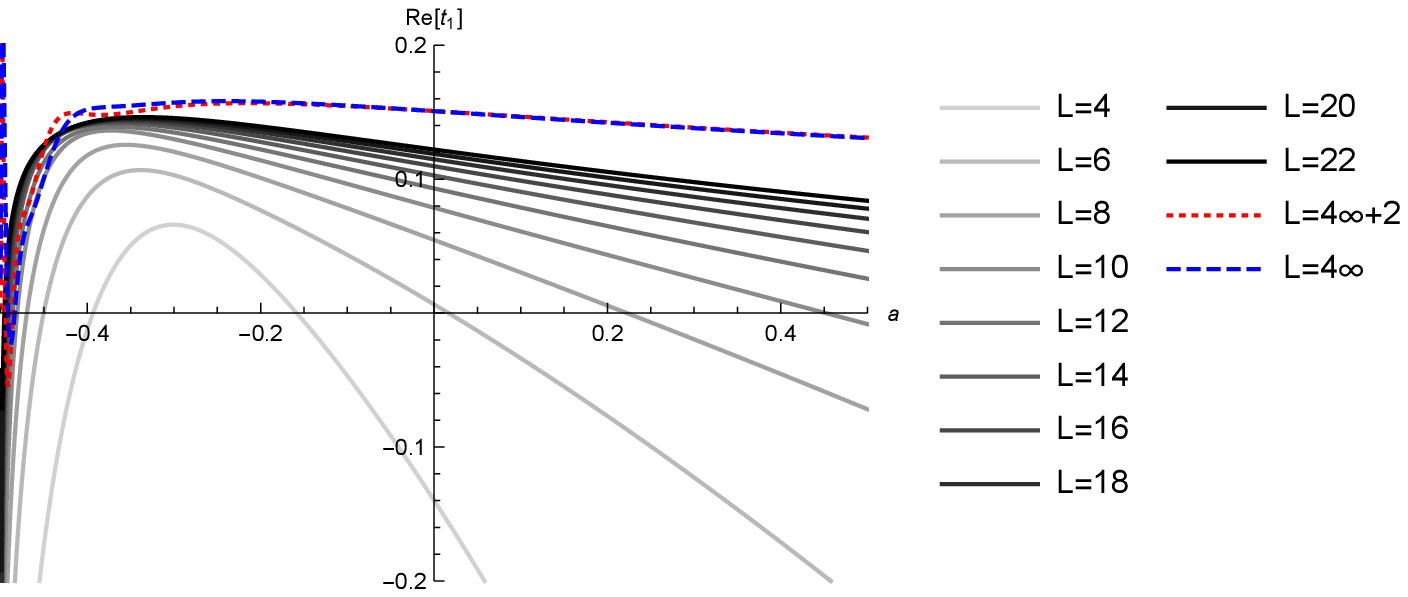}
 \caption{
 Plots of the real part of the coefficient of the lowest level state $c_1|0\rangle$
 for the ``ghost brane" solution $\Phi^{\rm G}_a$ at the truncation level $L$.
 The dotted and dashed lines are extrapolations to $L=4k+2$ ($k\to \infty$) and $L=4k$ ($k\to \infty$),
 respectively.
 The horizontal axis denotes the values of the parameter $a$.
  \label{fig:re_t1_gsol}
 }
 \end{figure}

\begin{figure}[htbp]
 \centering
 \includegraphics[height=6.5cm]{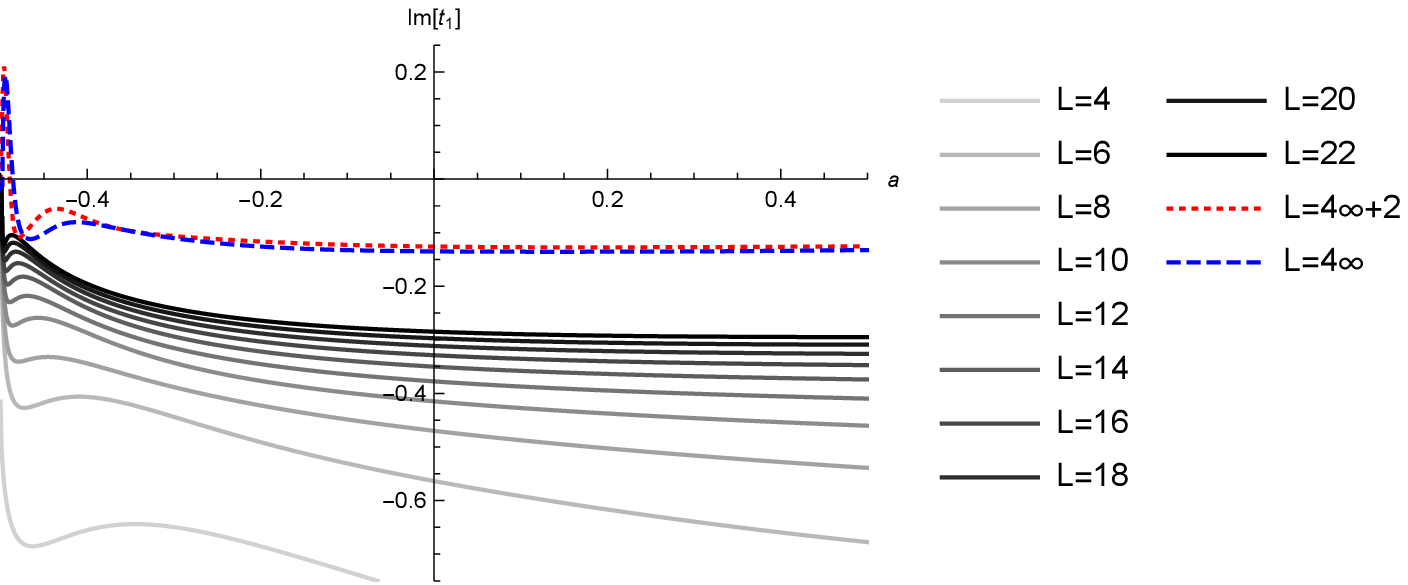}
 \caption{
 Plots of the imaginary part of the coefficient of the lowest level state $c_1|0\rangle$
 for the ``ghost brane" solution $\Phi^{\rm G}_a$ at the truncation level $L$.
 The dotted and dashed lines are extrapolations to $L=4k+2$ ($k\to \infty$) and $L=4k$ ($k\to \infty$),
 respectively.
 The horizontal axis denotes the values of the parameter $a$.
  \label{fig:im_t1_gsol}
 }
 \end{figure}

We list numerical data of three component fields for the solutions,
 $\Phi_a^{\rm D}$ and $\Phi_a^{\rm G}$, at $a=-1/2$ in 
 Tables \ref{tab:a=m1o2Dconfig} and  \ref{tab:a=m1o2Gconfig}.
 They can be compared with those of the solutions at $a=0$ listed in \cite{Kudrna:2018mxa}.
 Furthermore, we plot the coefficient of the lowest level state $c_1|0\rangle$, 
 or the tachyon field as a component, of $\Phi_a^{\rm D}$ and $\Phi_a^{\rm G}$ for $a\ge -1/2$
 in Figs.~\ref{fig:re_t1_dsol}, \ref{fig:im_t1_dsol}, \ref{fig:re_t1_gsol}, and \ref{fig:im_t1_gsol}.
  The data in the column of $c_1|0\rangle$ in Table \ref{tab:a=m1o2Dconfig} 
  correspond to points  at $a=-1/2$ in  Figs.~\ref{fig:re_t1_dsol} and  \ref{fig:im_t1_dsol}.
  The data in the column of $c_1|0\rangle$ in Table \ref{tab:a=m1o2Gconfig}
   correspond to points  at $a=-1/2$ in  Figs.~\ref{fig:re_t1_gsol} and  \ref{fig:im_t1_gsol}.
  
   Roughly, from Fig.~\ref{fig:im_t1_dsol},
 we can see that the imaginary part of the tachyon field  of  $\Phi_a^{\rm D}$ 
 vanishes with increasing level for $a\ge -1/2$.
 On the other hand, from Fig.~\ref{fig:im_t1_gsol},
 it seems that  the imaginary part of the tachyon field  of  $\Phi_a^{\rm G}$ 
 remains nonzero with increasing level for $a\gtrsim -0.4$.

\begin{table}[htbp]
        \caption{
 Coefficients of three lowest level states for the ``double brane" solution $\Phi^{\rm D}_{a=-1/2}$ at the truncation level $L$.
            \label{tab:a=m1o2Dconfig}}
    \begin{center}
%  %   \renewcommand{\arraystretch}{1.3}
\begin{tabular}{|r|lll|}
\hline
$L$&$c_1|0\rangle$&$L^{\rm mat}_{-2}c_1|0\rangle$&$L^{{\rm gh}\prime}_{-2}c_1|0\rangle$\\
\hline
2&$-1.62421 - 0.556839 i$
&$-0.144175 - 0.362514 i$
&$0.179079 - 0.141459 i$
\\
%%\hline
4&$-0.811826 - 0.492565 i$
&$0.0805148 - 0.211769 i$
&$0.0936975 - 0.0673882 i$
\\
%%\hline
6&$-0.617433 - 0.708114 i$
&$0.169208 - 0.223221 i$
&$0.0939303 - 0.0907553 i$
\\
%%\hline
8&$-0.46957 - 0.539068 i$
&$0.190683 - 0.158602 i$
&$0.127765 - 0.0634097 i$
\\
%%\hline
10&$-0.594608 - 0.464836 i$
&$0.110538 - 0.147441 i$
&$0.0882257 - 0.0634936 i$
\\
%%\hline
12&$-0.603815 - 0.328377 i$
&$0.0657693 - 0.115888 i$
&$0.084529 - 0.0536261 i$
\\
%%\hline
14&$-0.614146 - 0.359062 i$
&$0.0612609 - 0.120976 i$
&$0.0742589 - 0.05115 i$
\\
%%\hline
16&$-0.598281 - 0.280436 i$
&$0.0377076 - 0.0991098 i$
&$0.0693785 - 0.0454664 i$
\\
%%\hline
18&$-0.611011 - 0.304459 i$
&$0.0339221 - 0.106696 i$
&$0.0626943 - 0.045099 i$
\\
%%\hline
20&$-0.589014 - 0.25289 i$
&$0.0208091 - 0.0900697 i$
&$0.058484 - 0.0407456 i$
\\
%%\hline
22&$-0.601353 - 0.269899 i$
&$0.0173314 - 0.0970247 i$
&$0.0539424 - 0.0413202 i$
\\
\hline
\end{tabular}
% %      \renewcommand{\arraystretch}{1}
    \end{center}
\end{table}

%%%%%%%%
%%%%%%%%
%%%%%%%%

\begin{table}[htbp]
        \caption{
 Coefficients of three lowest level states for the ``ghost brane" solution $\Phi^{\rm G}_{a=-1/2}$ at the truncation level $L$.
            \label{tab:a=m1o2Gconfig}}
    \begin{center}
%  %   \renewcommand{\arraystretch}{1.3}
\begin{tabular}{|r|lll|}
\hline
$L$&$c_1|0\rangle$&$L^{\rm mat}_{-2}c_1|0\rangle$&$L^{{\rm gh}\prime}_{-2}c_1|0\rangle$\\
\hline
4&$-0.761439 - 0.415565 i$
&$-0.0430207 - 0.0388406 i$
&$1.44204 + 0.0481819 i$
\\
%%\hline
6&$-0.360559$
&$-0.0188028$
&$0.675305$
\\
%%\hline
8&$-0.238314$
&$-0.0162807$
&$0.440174$
\\
%%\hline
10&$-0.152385$
&$-0.0120697$
&$0.310303$
\\
%%\hline
12&$-0.11406$
&$-0.0102722$
&$0.234504$
\\
%%\hline
14&$-0.0876269$
&$-0.00885158$
&$0.184856$
\\
%%\hline
16&$-0.0708067$
&$-0.00777904$
&$0.150024$
\\
%%\hline
18&$-0.0584666$
&$-0.00695981$
&$0.124815$
\\
%%\hline
20&$-0.0494588$
&$-0.00625865$
&$0.105564$
\\
%%\hline
22&$-0.0425024$
&$-0.00570499$
&$0.0907437$
\\
\hline
\end{tabular}
% %      \renewcommand{\arraystretch}{1}
    \end{center}
\end{table}

\subsection{Evaluation of quadratic identities}

It is known that solutions to the equation of motion in open string field theory satisfy certain 
quadratic identities \cite{Schnabl:2000wt}.
Here, we show some numerical data for evaluation of them as a consistency check 
of numerical solutions in the theory around the TT solution with the parameter $a$.
We can derive quadratic identities for solutions to (\ref{eq:EOMQp}):
\begin{align}
\langle \Phi,[Q^{\prime},L^{\rm mat}_n]\Phi\rangle
&=-\dfrac{65}{54}(-1)^{\frac{n}{2}}n\,\delta_{n:\,\text{even}}\langle \Phi,Q^{\prime}\Phi\rangle,\\
\langle \Phi,[Q^{\prime},L_n^{\prime\,{\rm tot}}]\Phi\rangle
&=-(-1)^{\frac{n}{2}}n\,\delta_{n:\,\text{even}}\langle \Phi,Q^{\prime}\Phi\rangle,
\end{align}
where $L_n^{\prime{\rm tot}}=L_n^{\rm mat}+L_n^{\prime\,{\rm gh}}$,
using symmetry of the interaction term of  the action (\ref{eq:S_aPhi}).
Particularly, in the case of Siegel gauge solution $\Phi$, if $\langle \Phi,c_0L(a)\Phi\rangle\ne 0$, 
the ratios of the left hand side and the right hand side of the above equations with $n=2m$:
\begin{align}
R_m&=
\begin{cases}
\dfrac{\langle\Phi,c_0\left((1+a)L_{2m}^{\rm mat}+a\frac{m-1}{2m}L_{2m+2}^{\rm mat}
+a\frac{m+1}{2m}L_{2m-2}^{\rm mat}\right)\Phi\rangle}
{(-1)^m\frac{65}{54}\langle \Phi,c_0\left((1+a)(L_0^{\prime\,{\rm tot}}-1)+aL_{2}^{\prime\,{\rm tot}}+4aZ(a)\right)\Phi\rangle}&
(m=2,3,4,\cdots)\\
\dfrac{-\langle\Phi,c_0\left((1+a)L_{2}^{\rm mat}
+aL_0^{\rm mat}+\frac{13}{4}a\right)\Phi\rangle}
{\frac{65}{54}\langle \Phi,c_0\left((1+a)(L_0^{\prime\,{\rm tot}}-1)+aL_{2}^{\prime\,{\rm tot}}+4aZ(a)\right)\Phi\rangle}&(m=1)
\end{cases},
\label{eq:Rmprime}\\
\tilde R_m&=\begin{cases}
\dfrac{\langle\Phi,c_0\left((1+a)L_{2m}^{\prime\,{\rm tot}}
+a\frac{m-1}{2m}L_{2m+2}^{\prime\,{\rm tot}}
+a\frac{m+1}{2m}L_{2m-2}^{\prime\,{\rm tot}}\right)\Phi\rangle}
{(-1)^m
\langle \Phi,c_0\left((1+a)(L_0^{\prime\,{\rm tot}}-1)+aL_{2}^{\prime\,{\rm tot}}+4aZ(a)\right)\Phi\rangle}&(m=2,3,4\cdots)\\
\dfrac{-\langle\Phi,c_0\left((1+a)L_{2}^{\prime\,{\rm tot}}
+a(L_0^{\prime\,{\rm tot}}+3)\right)\Phi\rangle}
{\langle \Phi,c_0\left((1+a)(L_0^{\prime\,{\rm tot}}-1)+aL_{2}^{\prime\,{\rm tot}}+4aZ(a)\right)\Phi\rangle}&(m=1)
\end{cases}
\label{eq:tildeRmprime}
\end{align}
should be one identically\footnote{
The details can be found in \cite{Kishimoto:2021}, where numerical data 
for the tachyon vacuum and single brane solutions  at $a=-1/2$,
based on the solutions in \cite{Kishimoto:2011zza},
are listed up to level $26$.
}. 
We note that $\tilde R_1$ is one at $a=-1/2$ trivially.

\begin{figure}[htbp]
 \centering
 \includegraphics[height=6.5cm]{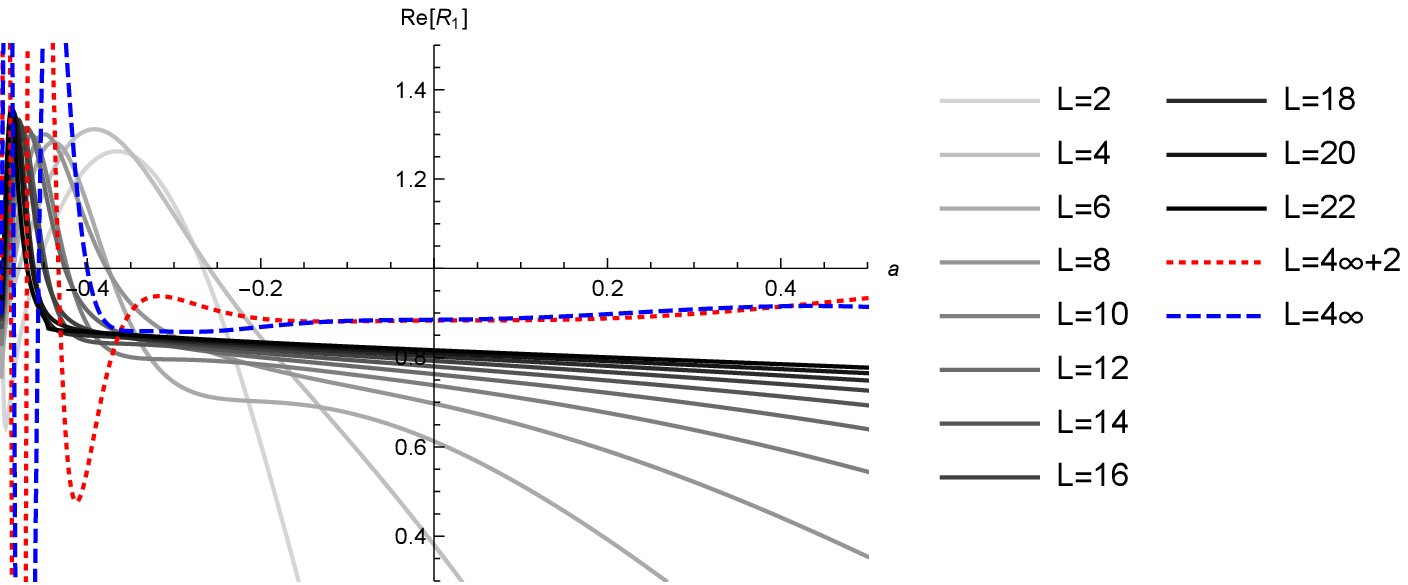}
 \caption{
 Plots of the real part of $R_1$ (\ref{eq:Rmprime}) of $\Phi_a^{\rm D}$
 at the truncation level $L$.
 The dotted and dashed lines are extrapolations to $L=4k+2$ and $L=4k$ ($k\to \infty$),
 respectively.
 The horizontal axis denotes the value of the parameter $a$ at ${\rm Re}\,R_1=1$.
  \label{fig:re_r1_dsol}
 }
 \end{figure}

\begin{figure}[htbp]
 \centering
 \includegraphics[height=6.5cm]{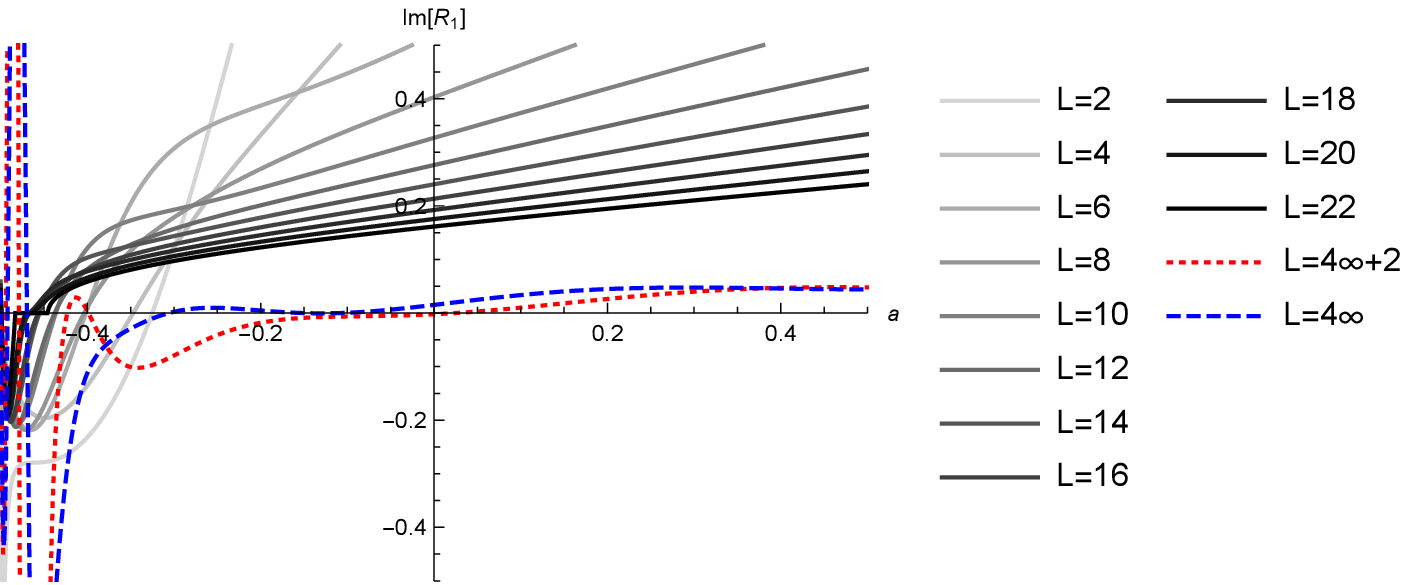}
 \caption{
 Plots of the imaginary part of $R_1$ (\ref{eq:Rmprime}) of $\Phi_a^{\rm D}$
 at the truncation level $L$.
 The dotted and dashed lines are extrapolations to $L=4k+2$ ($k\to \infty$) and $L=4k$ ($k\to \infty$),
 respectively.
 The horizontal axis denotes the value of the parameter $a$.
  \label{fig:im_r1_dsol}
 }
 \end{figure}

\begin{figure}[htbp]
 \centering
 \includegraphics[height=6.5cm]{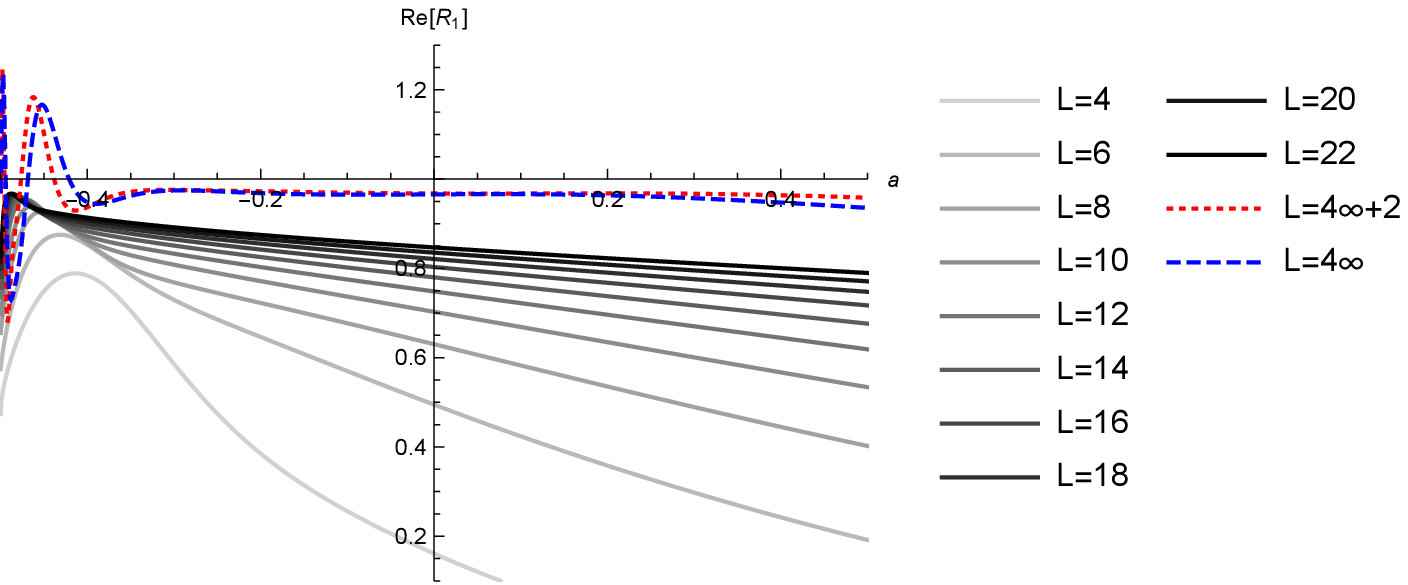}
 \caption{
 Plots of the real part of $R_1$ (\ref{eq:Rmprime}) of $\Phi_a^{\rm G}$
 at the truncation level $L$.
 The dotted and dashed lines are extrapolations to $L=4k+2$ ($k\to \infty$) and $L=4k$ ($k\to \infty$),
 respectively.
 The horizontal axis denotes the value of the parameter $a$ at ${\rm Re}\,R_1=1$.
  \label{fig:re_r1_gsol}
 }
 \end{figure}

\begin{figure}[htbp]
 \centering
 \includegraphics[height=6.5cm]{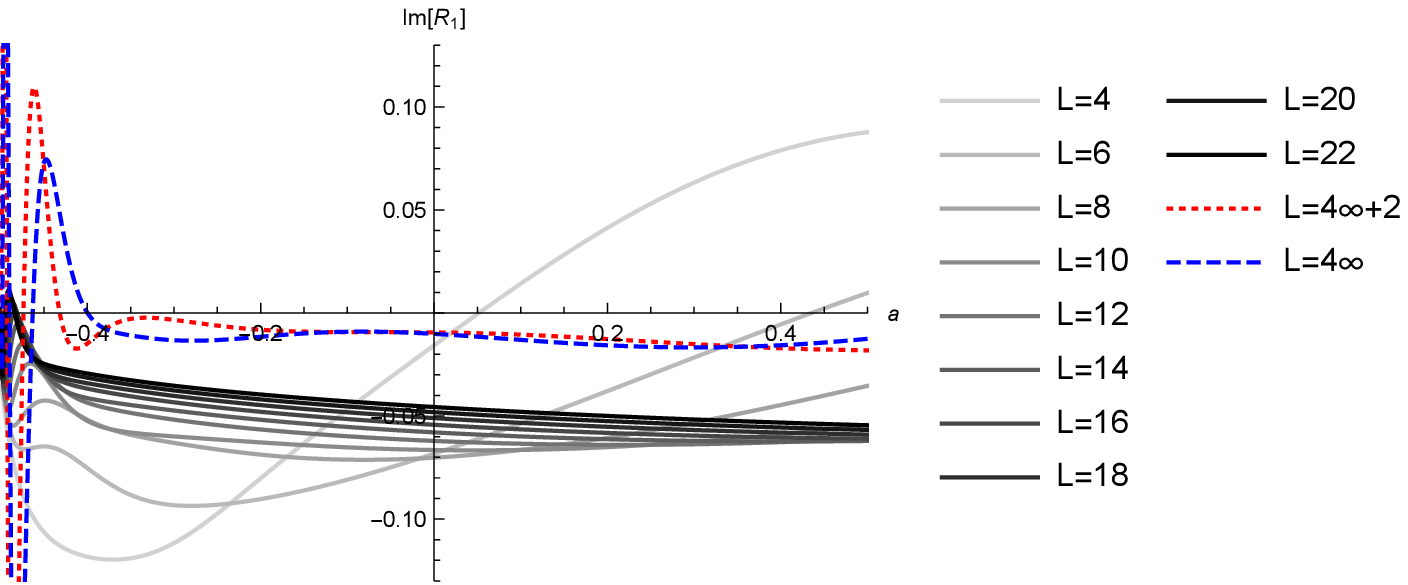}
 \caption{
  Plots of the imaginary part of $R_1$ (\ref{eq:Rmprime}) of $\Phi_a^{\rm G}$
 at the truncation level $L$.
 The dotted and dashed lines are extrapolations to $L=4k+2$ ($k\to \infty$) and $L=4k$ ($k\to \infty$),
 respectively.
 The horizontal axis denotes the value of the parameter $a$.
  \label{fig:im_r1_gsol}
 }
 \end{figure}

We list some numerical data for the evaluations of 
$R_m$ (\ref{eq:Rmprime}) and $\tilde R_m$  (\ref{eq:tildeRmprime})
 of the ``double brane" solution at $a=-1/2$ in Tables \ref{tab:KTDRm1} and \ref{tab:KTDtRm1}
and those of the ``ghost brane" solution at $a=-1/2$
 in Tables \ref{tab:GRm1} and \ref{tab:GtRm1}.
 We also plot the value of $R_1$ of $\Phi_a^{\rm D}$ in Figs. \ref{fig:re_r1_dsol} and \ref{fig:im_r1_dsol}
 and that of $\Phi_a^{\rm G}$ in Figs. \ref{fig:re_r1_gsol} and \ref{fig:im_r1_gsol}.
 Roughly, from these Figures, it seems that $R_1$ for $\Phi_a^{\rm D}$ and $\Phi_a^{\rm G}$
 approaches one with increasing level although it is unstable around $a=-1/2$.
 
 The above numerical behavior is consistent with quadratic identities
 $R_m=1$ and $\tilde R_m=1$ for solutions at $L=\infty$.
\begin{table}[htbp]
\caption{$R_m$ (\ref{eq:Rmprime}) of $\Phi_{a=-1/2}^{\rm D}$
at the truncation level $L$
for $m=1,2,3,4$.
\label{tab:KTDRm1}
}
\centering
{\small
\begin{tabular}{|r| l l l l|}
\hline
$L$&
$R_1$&
$R_2$&
$R_3$&
$R_4$
\\
\hline
$2$&
$0.911488 - 0.152245 i$&
$-1.27584 + 0.412182 i$&
$0$&
$0$
\\
%\hline
$4$&
$0.885585 + 0.021908 i$&
$-1.28084 - 0.320084 i$&
$-1.2104 - 0.84751 i$&
$0$
\\
%\hline
$6$&
$0.8729 + 0.0201808 i$&
$-1.60019 + 1.10353 i$&
$-1.59739 + 1.49291 i$&
$-0.961762 + 0.978238 i$
\\
%\hline
$8$&
$0.950585 + 0.00241884 i$&
$2.06239 + 0.84862 i$&
$3.66403 - 2.82962 i$&
$-2.57524 - 3.66336 i$
\\
%\hline
$10$&
$0.893514 + 0.0530564 i$&
$-0.248418 + 1.30193 i$&
$0.9719 + 1.85136 i$&
$-1.11547 + 2.07613 i$
\\
%\hline
$12$&
$0.937224 + 0.0359425 i$&
$0.461938 + 0.777509 i$&
$1.56888 + 0.298369 i$&
$-0.28761 + 0.782399 i$
\\
%\hline
$14$&
$0.929671 + 0.058921 i$&
$0.597104 + 0.95585 i$&
$1.36524 + 0.815189 i$&
$-0.0304565 + 1.67951 i$
\\
%\hline
$16$&
$0.953995 + 0.036186 i$&
$0.765202 + 0.59013 i$&
$1.27848 + 0.184195 i$&
$0.324349 + 0.7959 i$
\\
%\hline
$18$&
$0.953387 + 0.0515589 i$&
$0.878275 + 0.620112 i$&
$1.2123 + 0.34281 i$&
$0.606532 + 1.09488 i$
\\
%\hline
$20$&
$0.96477 + 0.0330054 i$&
$0.890854 + 0.427151 i$&
$1.12715 + 0.112292 i$&
$0.678698 + 0.623308 i$
\\
%\hline
$22$&
$0.966273 + 0.0431989 i$&
$0.959863 + 0.420163 i$&
$1.10173 + 0.175657 i$&
$0.858197 + 0.707533 i$
\\
\hline
\end{tabular}
}
\end{table}

\begin{table}[htbp]
\caption{$\tilde R_m$ (\ref{eq:tildeRmprime}) of $\Phi_{a=-1/2}^{\rm D}$
at the truncation level $L$
for $m=1,2,3,4$.
\label{tab:KTDtRm1}
}
\centering
{\small
\begin{tabular}{|r|l l l l|}
\hline
$L$&
$\tilde R_1$\qquad~&
$\tilde R_2$&
$\tilde R_3$&
$\tilde R_4$
\rule[0pt]{0pt}{1.1em}
\\
\hline
$2$&
$1$&
$-1.49458 + 0.555977 i$&
$0$&
$0$
\\
%\hline
$4$&
$1$&
$-1.3959 - 0.372793 i$&
$-1.32557 - 0.977179 i$&
$0$
\\
%\hline
$6$&
$1$&
$-1.82477 + 1.15122 i$&
$-1.89735 + 1.67941 i$&
$-1.1746 + 1.14418 i$
\\
%\hline
$8$&
$1$&
$2.26372 + 0.934943 i$&
$4.01471 - 3.30933 i$&
$-2.90247 - 4.17631 i$
\\
%\hline
$10$&
$1$&
$-0.344922 + 1.40266 i$&
$1.06346 + 2.07209 i$&
$-1.32026 + 2.17119 i$
\\
%\hline
$12$&
$1$&
$0.467894 + 0.820665 i$&
$1.71676 + 0.257096 i$&
$-0.355223 + 0.778747 i$
\\
%\hline
$14$&
$1$&
$0.596248 + 1.02744 i$&
$1.5074 + 0.860021 i$&
$-0.143872 + 1.76552 i$
\\
%\hline
$16$&
$1$&
$0.786884 + 0.614174 i$&
$1.36667 + 0.142577 i$&
$0.294999 + 0.813055 i$
\\
%\hline
$18$&
$1$&
$0.901552 + 0.656254 i$&
$1.30334 + 0.320084 i$&
$0.573704 + 1.15352 i$
\\
%\hline
$20$&
$1$&
$0.914962 + 0.437798 i$&
$1.18301 + 0.0740143 i$&
$0.676277 + 0.637951 i$
\\
%\hline
$22$&
$1$&
$0.9852 + 0.436868 i$&
$1.15838 + 0.139992 i$&
$0.859524 + 0.740684 i$
\\
\hline
\end{tabular}
}
\end{table}

%%%%%%%%%%%%
%%%%%%%%%%%%

\begin{table}[htbp]
\caption{$R_m$ (\ref{eq:Rmprime}) of $\Phi_{a=-1/2}^{\rm G}$
at the truncation level $L$
for $m=1,2,3,4$.
\label{tab:GRm1}
}
\centering
{\small
\begin{tabular}{|r| l l l l|}
\hline
$L$&
$R_1$&
$R_2$&
$R_3$&
$R_4$
\\
\hline
$4$&
$0.473669 - 0.0224943 i$&
$0.0290573 - 0.0136117 i$&
$-0.00391726 - 0.0214949 i$&
$0$
\\
%\hline
$6$&
$0.574357$&
$0.142781$&
$0.0199572$&
$-0.00288513$
\\
%\hline
$8$&
$0.6554$&
$0.340813$&
$0.152805$&
$0.0118893$
\\
%\hline
$10$&
$0.698864$&
$0.479036$&
$0.328676$&
$0.100043$
\\
%\hline
$12$&
$0.736879$&
$0.566131$&
$0.462773$&
$0.244439$
\\
%\hline
$14$&
$0.761953$&
$0.631667$&
$0.555784$&
$0.373664$
\\
%\hline
$16$&
$0.78442$&
$0.677713$&
$0.625193$&
$0.473761$
\\
%\hline
$18$&
$0.800922$&
$0.714594$&
$0.674924$&
$0.551411$
\\
%\hline
$20$&
$0.815886$&
$0.742564$&
$0.714176$&
$0.610725$
\\
%\hline
$22$&
$0.827609$&
$0.765937$&
$0.743846$&
$0.657547$
\\
\hline
\end{tabular}
}
\end{table}

\begin{table}[htbp]
\caption{$\tilde R_m$ (\ref{eq:tildeRmprime}) of $\Phi_{a=-1/2}^{\rm G}$
at the truncation level $L$
for $m=1,2,3,4$.
\label{tab:GtRm1}
}
\centering
{\small
\begin{tabular}{|r| l l l l|}
\hline
$L$&
$\tilde R_1$\qquad~&
$\tilde R_2$&
$\tilde R_3$&
$\tilde R_4$
\rule[0pt]{0pt}{1.1em}
\\
\hline
$4$&
$1$&
$-0.0219749 + 0.000400093 i$&
$-0.000350886 - 0.0220953 i$&
$0$
\\
%\hline
$6$&
$1$&
$-0.0293363$&
$0.029805$&
$0.00101128$
\\
%\hline
$8$&
$1$&
$0.16492$&
$0.215592$&
$0.00493668$
\\
%\hline
$10$&
$1$&
$0.316654$&
$0.431604$&
$0.0656148$
\\
%\hline
$12$&
$1$&
$0.411717$&
$0.583257$&
$0.197593$
\\
%\hline
$14$&
$1$&
$0.490697$&
$0.68693$&
$0.318767$
\\
%\hline
$16$&
$1$&
$0.545656$&
$0.758987$&
$0.413544$
\\
%\hline
$18$&
$1$&
$0.592945$&
$0.809749$&
$0.489035$
\\
%\hline
$20$&
$1$&
$0.62858$&
$0.847047$&
$0.546827$
\\
%\hline
$22$&
$1$&
$0.659964$&
$0.874591$&
$0.593859$
\\
\hline
\end{tabular}
}
\end{table}

%%%%%%%%%%%%%%%%%%%%%%%%%%%%%%%%%%%%%%%%%%%%%%%%%%%%%
\bibliographystyle{utphys}
\bibliography{referenceik}

\providecommand{\href}[2]{#2}\begingroup\raggedright\begin{thebibliography}{10}

\bibitem{Sen:1999nx}
A.~Sen and B.~Zwiebach, ``{Tachyon condensation in string field theory},''
  \href{http://dx.doi.org/10.1088/1126-6708/2000/03/002}{{\em JHEP} {\bf 03}
  (2000)  002}, \href{http://arxiv.org/abs/hep-th/9912249}{{\tt
  arXiv:hep-th/9912249}}.

\bibitem{Moeller:2000xv}
N.~Moeller and W.~Taylor, ``{Level truncation and the tachyon in open bosonic
  string field theory},''
  \href{http://dx.doi.org/10.1016/S0550-3213(00)00293-5}{{\em Nucl. Phys. B}
  {\bf 583} (2000)  105--144}, \href{http://arxiv.org/abs/hep-th/0002237}{{\tt
  arXiv:hep-th/0002237}}.

\bibitem{Gaiotto:2002wy}
D.~Gaiotto and L.~Rastelli, ``{Experimental string field theory},''
  \href{http://dx.doi.org/10.1088/1126-6708/2003/08/048}{{\em JHEP} {\bf 08}
  (2003)  048}, \href{http://arxiv.org/abs/hep-th/0211012}{{\tt
  arXiv:hep-th/0211012}}.

\bibitem{Kishimoto:2011zza}
I.~Kishimoto, ``{On numerical solutions in open string field theory},''
  \href{http://dx.doi.org/10.1143/PTPS.188.155}{{\em Prog. Theor. Phys. Suppl.}
  {\bf 188} (2011)  155--162}.

\bibitem{Kudrna:2018mxa}
M.~Kudrna and M.~Schnabl, ``{Universal Solutions in Open String Field
  Theory},'' \href{http://arxiv.org/abs/1812.03221}{{\tt arXiv:1812.03221
  [hep-th]}}.

\bibitem{Takahashi:2002ez}
T.~Takahashi and S.~Tanimoto, ``{Marginal and scalar solutions in cubic open
  string field theory},''
  \href{http://dx.doi.org/10.1088/1126-6708/2002/03/033}{{\em JHEP} {\bf 03}
  (2002)  033}, \href{http://arxiv.org/abs/hep-th/0202133}{{\tt
  arXiv:hep-th/0202133}}.

\bibitem{Kishimoto:2002xi}
I.~Kishimoto and T.~Takahashi, ``{Open string field theory around universal
  solutions},'' \href{http://dx.doi.org/10.1143/PTP.108.591}{{\em Prog. Theor.
  Phys.} {\bf 108} (2002)  591--602},
  \href{http://arxiv.org/abs/hep-th/0205275}{{\tt arXiv:hep-th/0205275}}.

\bibitem{Takahashi:2003ppa}
T.~Takahashi, ``{Tachyon condensation and universal solutions in string field
  theory},'' \href{http://dx.doi.org/10.1016/j.nuclphysb.2003.08.007}{{\em
  Nucl. Phys. B} {\bf 670} (2003)  161--182},
  \href{http://arxiv.org/abs/hep-th/0302182}{{\tt arXiv:hep-th/0302182}}.

\bibitem{Kishimoto:2009nd}
I.~Kishimoto and T.~Takahashi, ``{Vacuum structure around identity based
  solutions},'' \href{http://dx.doi.org/10.1143/PTP.122.385}{{\em Prog. Theor.
  Phys.} {\bf 122} (2009)  385--399},
  \href{http://arxiv.org/abs/0904.1095}{{\tt arXiv:0904.1095 [hep-th]}}.

\bibitem{Kishimoto:2009hc}
I.~Kishimoto and T.~Takahashi, ``{Exploring Vacuum Structure around
  Identity-Based Solutions},''
  \href{http://dx.doi.org/10.1007/s11232-010-0055-x}{{\em Theor. Math. Phys.}
  {\bf 163} (2010)  717--724}, \href{http://arxiv.org/abs/0910.3026}{{\tt
  arXiv:0910.3026 [hep-th]}}.

\bibitem{Ishibashi:2014mua}
N.~Ishibashi, ``{Comments on Takahashi-Tanimoto\textquoteright{}s scalar
  solution},'' \href{http://dx.doi.org/10.1007/JHEP02(2015)168}{{\em JHEP} {\bf
  02} (2015)  168}, \href{http://arxiv.org/abs/1408.6319}{{\tt arXiv:1408.6319
  [hep-th]}}.

\bibitem{Kishimoto:2014lua}
I.~Kishimoto, T.~Masuda, and T.~Takahashi, ``{Observables for identity-based
  tachyon vacuum solutions},''
  \href{http://dx.doi.org/10.1093/ptep/ptu136}{{\em PTEP} {\bf 2014} (2014)
  no.~10, 103B02}, \href{http://arxiv.org/abs/1408.6318}{{\tt arXiv:1408.6318
  [hep-th]}}.

\bibitem{Kawano:2008ry}
T.~Kawano, I.~Kishimoto, and T.~Takahashi, ``{Gauge Invariant Overlaps for
  Classical Solutions in Open String Field Theory},''
  \href{http://dx.doi.org/10.1016/j.nuclphysb.2008.05.025}{{\em Nucl. Phys. B}
  {\bf 803} (2008)  135--165}, \href{http://arxiv.org/abs/0804.1541}{{\tt
  arXiv:0804.1541 [hep-th]}}.

\bibitem{Ellwood:2008jh}
I.~Ellwood, ``{The Closed string tadpole in open string field theory},''
  \href{http://dx.doi.org/10.1088/1126-6708/2008/08/063}{{\em JHEP} {\bf 08}
  (2008)  063}, \href{http://arxiv.org/abs/0804.1131}{{\tt arXiv:0804.1131
  [hep-th]}}.

\bibitem{Schnabl:2005gv}
M.~Schnabl, ``{Analytic solution for tachyon condensation in open string field
  theory},'' \href{http://dx.doi.org/10.4310/ATMP.2006.v10.n4.a1}{{\em Adv.
  Theor. Math. Phys.} {\bf 10} (2006) no.~4, 433--501},
  \href{http://arxiv.org/abs/hep-th/0511286}{{\tt arXiv:hep-th/0511286}}.

\bibitem{Hata:2000bj}
H.~Hata and S.~Shinohara, ``{BRST invariance of the nonperturbative vacuum in
  bosonic open string field theory},''
  \href{http://dx.doi.org/10.1088/1126-6708/2000/09/035}{{\em JHEP} {\bf 09}
  (2000)  035}, \href{http://arxiv.org/abs/hep-th/0009105}{{\tt
  arXiv:hep-th/0009105}}.

\bibitem{Baba:2012cs}
T.~Baba and N.~Ishibashi, ``{Energy from the gauge invariant observables},''
  \href{http://dx.doi.org/10.1007/JHEP04(2013)050}{{\em JHEP} {\bf 04} (2013)
  050}, \href{http://arxiv.org/abs/1208.6206}{{\tt arXiv:1208.6206 [hep-th]}}.

\bibitem{Asano:2006hk}
M.~Asano and M.~Kato, ``{New Covariant Gauges in String Field Theory},''
  \href{http://dx.doi.org/10.1143/PTP.117.569}{{\em Prog. Theor. Phys.} {\bf
  117} (2007)  569--587}, \href{http://arxiv.org/abs/hep-th/0611189}{{\tt
  arXiv:hep-th/0611189}}.

\bibitem{Kishimoto:2009cz}
I.~Kishimoto and T.~Takahashi, ``{Numerical Evaluation of Gauge Invariants for
  $a$-gauge Solutions in Open String Field Theory},''
  \href{http://dx.doi.org/10.1143/PTP.121.695}{{\em Prog. Theor. Phys.} {\bf
  121} (2009)  695--710}, \href{http://arxiv.org/abs/0902.0445}{{\tt
  arXiv:0902.0445 [hep-th]}}.

\bibitem{Kishimoto:2009hb}
I.~Kishimoto and T.~Takahashi, ``{Numerical Evaluation of Gauge Invariants for
  a-gauge Solutions in Open String Field Theory},''
  \href{http://dx.doi.org/10.1007/s11232-010-0054-y}{{\em Theor. Math. Phys.}
  {\bf 163} (2010)  710--716}, \href{http://arxiv.org/abs/0910.3025}{{\tt
  arXiv:0910.3025 [hep-th]}}.

\bibitem{Arroyo:2019liw}
E.~Aldo~Arroyo and M.~Kudrna, ``{Numerical solution for tachyon vacuum in the
  Schnabl gauge},'' \href{http://dx.doi.org/10.1007/JHEP02(2020)065}{{\em JHEP}
  {\bf 02} (2020)  065}, \href{http://arxiv.org/abs/1908.05330}{{\tt
  arXiv:1908.05330 [hep-th]}}.

\bibitem{Schnabl:2000wt}
M.~Schnabl, ``{Constraints on the tachyon condensate from anomalous
  symmetries},'' \href{http://dx.doi.org/10.1016/S0370-2693(01)00282-9}{{\em
  Phys. Lett. B} {\bf 504} (2001)  61--63},
  \href{http://arxiv.org/abs/hep-th/0011238}{{\tt arXiv:hep-th/0011238}}.

\bibitem{Kishimoto:2021}
I.~Kishimoto, ``{Numerical evaluation of quadratic identities for classical
  solutions in open string field theory (in Japanese)},'' {\em Studies in
  Liberal Arts and Sciences, Tokyo University of Science} {\bf 53} (2021)  (to
  be published).

\end{thebibliography}\endgroup

\end{document}